\documentclass[journal]{IEEEtran}
\usepackage{rotating}
\usepackage{lscape}
\usepackage{longtable}
\usepackage[utf8]{inputenc}
\usepackage[dvipsnames]{xcolor}
\usepackage[section]{placeins}
\usepackage{adjustbox}
\usepackage{array}
\usepackage{tabularx,ragged2e,booktabs}
\usepackage{tabulary,booktabs}
\usepackage{ragged2e}
\usepackage{booktabs,chemformula}
\usepackage{nidanfloat}
\usepackage[numbers]{natbib}
\usepackage{comment}
\usepackage{graphicx}
\usepackage{rotating}
\usepackage{lscape}
\usepackage{longtable}
\usepackage[utf8]{inputenc}
\usepackage[dvipsnames]{xcolor}
\usepackage[section]{placeins}
\usepackage{adjustbox}
\usepackage{array}
\usepackage{tabularx,ragged2e,booktabs}
\usepackage{tabulary,booktabs}
\usepackage{ragged2e}
\usepackage{booktabs,chemformula}
\usepackage{nidanfloat}
\usepackage[numbers]{natbib}
\usepackage{comment}
\usepackage{graphicx}
\usepackage{multirow}
\usepackage{neuralnetwork}
\usepackage{array,ragged2e}
\newcolumntype{P}[1]{>{\RaggedRight\arraybackslash}p{#1}}
\usepackage{blindtext}
\graphicspath{ {./images/} }
\usepackage{authblk}
\ifCLASSINFOpdf
\else
\fi
\begin{document}
\title{Adversarial Machine Learning \\In Network Intrusion Detection Domain: \\A Systematic Review}

\author[1, 2]{Huda Ali Alatwi\thanks{h.a.alatwi@newcastle.ac.uk}}
\author[1]{Charles Morisset\thanks{charles.morisset@newcastle.ac.uk}}

\affil[1]{Newcastle University, UK}
\affil[2]{Tabuk University, KSA}

\maketitle
\begin{abstract}
Due to their massive success in various domains, deep learning techniques are increasingly used to design network intrusion detection solutions that detect and mitigate unknown and known attacks with high accuracy detection rates and minimal feature engineering. However, it has been found that deep learning models are vulnerable to data instances that can mislead the model to make incorrect classification decisions so-called (adversarial examples). Such vulnerability allows attackers to target NIDSs by adding small crafty perturbations to the malicious traffic to evade detection and disrupt the system's critical functionalities. The problem of deep adversarial learning has been extensively studied in the computer vision domain; however, it is still an area of open research in network security applications. Therefore, this survey explores the researches that employ different aspects of adversarial machine learning in the area of network intrusion detection in order to provide directions for potential solutions. First, the surveyed studies are categorized based on their contribution to generating adversarial examples, evaluating the robustness of ML-based NIDs towards adversarial examples, and defending these models against such attacks. Second, we highlight the characteristics identified in the surveyed research. Furthermore, we discuss the applicability of the existing generic adversarial attacks for the NIDS domain, the feasibility of launching the proposed attacks in real-world scenarios, and the limitations of the existing mitigation solutions. 
\end{abstract}
\renewcommand\IEEEkeywordsname{Keywords}
\IEEEpeerreviewmaketitle
\section{Introduction}  
\IEEEPARstart{A} {s the} first line of defense, the network intrusion detection system (NIDS) carries the responsibility of monitoring network traffic to recognize any malicious and anomalous activities that can be part of an attack~\cite{buczak2015survey}. A NIDS can be signature-based by matching suspicious traffic to predefined signatures of known attacks based on (e.g., port number, IP address, protocol, etc.), or anomaly-based by measuring the variance between observed traffic and the trustworthy baseline traffic, thus potentially capturing zero-day attacks~\cite{mishra2018detailed}. 

A wide range of machine learning techniques has been investigated for security applications: malware detection~\cite{ali2016android}, spam detection~\cite{crawford2015survey}, network anomaly detection~\cite{mishra2018detailed}. Deep learning (DL) techniques, in particular, are getting considerable interest due to their capabilities of automatic feature engineering and self-learning, along with providing high accuracy rates. The reduced cost of parallel computation hardware has enabled different DL approaches to address many challenges in network intrusion detection domain~\cite{shone2018deep}. However, while machine learning-based security solutions address some issues, they also introduce new ones, such as adversarial machine learning attacks~\cite{huang2011adversarial}. It has been found that DL models are vulnerable to so-called adversarial examples (AEs), i.e., when a crafted data instance can mislead the model to make incorrect classification decisions~\cite{szegedy2013intriguing}. 
Adversarial DL has been extensively studied in computer vision and image recognition; however, it is still an area of open research in network security applications. The adversary can probe the NIDS in a black-box manner using crafted AEs by repeatedly modifying a small subset of the traffic features at each attack attempt and querying the detector for some feedback that indicates whether the attempt succeeds or fails (e.g., an acknowledgment or no response)~\cite{zhang2020tiki}. According to the received feedback, attackers can insert new values or tweak the perturbations to specifically selected features till succeeding in evading the NIDS~\cite{zhang2020tiki}. 

Adversarial attacks are a growing challenge, and ML models' protection and resilience against such attacks need to be addressed. Several works in the image and text recognition domains have investigated the threat of AEs and proposed possible mitigation solutions against them.
However, there is a lack of research addressing the challenges of adversarial attacks in the NIDS domain. This area of research is still in its early stages. Therefore, that serves as a motivation to introduce a comprehensive survey that establishes directions for further research. In contrast with a survey by~\citeauthor{corona2013adversarial}~\cite{corona2013adversarial} which focused on the weaknesses that can be exploited at different abstraction levels in the NIDS in general, the attacks that exploit vulnerabilities in the NIDS employ ML models as detection mechanisms are the main focus of this study. 

The contribution of this survey is as follows:
\begin{itemize}
\item to summarize and investigate the recent advances of adversarial learning applied to the NIDS domain. 
\item to outline the main challenges, and to identify open issues and certain areas that demand further research for considerable impacts.
\end{itemize}

To the best of our knowledge, there is no recent research work has focused mainly on providing a comprehensive review of the adversarial machine learning attacks against deep learning applications in the context of the network anomaly detection domain. 

The rest of this survey is structured as follows: in Sec. ~\ref{relatedwork}, previous related surveys are discussed. Sec.~\ref{literaturereview} describes the researches that applied adversarial machine learning into the NIDS domain. In Sec.~\ref{discussion}, an overall discussion of the surveyed studies, our remarkable findings, guidance for future works, and research opportunities are provided. Lastly, Sec.~\ref{conclusion} provides a conclusion for our study.

 \label{introduction}
\section{Related Work}
 This review is an extension of a previously published paper~\cite{alatwi2021adversarial} that mainly focused on adversarial black-box attacks.  This extended paper additionally covers white-box attacks, provides in-depth analysis for our findings, addresses the gaps, and sets directions for further research. In other related works,  ~\citeauthor{buczak2015survey}~\cite{buczak2015survey} discussed the complexity and challenges of using machine learning and data mining approaches for network intrusion detection; however, the threats of adversarial attacks were outside the scope of their study.~\citeauthor{biggio2018wild}~\cite{biggio2018wild} provided a detailed overview on the evolution of adversarial learning in the domains of computer vision and cybersecurity; nevertheless, this problem in the area of network anomaly detection was not covered.~\citeauthor{qiu2019review}~\cite{qiu2019review} presented a general review on adversarial machine learning attacks against a wide range of deep learning-based AI applications with a concise consideration of some security applications such as intrusion detection, malware detection, and cloud security.~\citeauthor{ibitoye2019threat}~\cite{ibitoye2019threat} provided a survey on some of the adversarial attacks against machine learning applications in network security, including intrusion detection, spam filtering, and malware detection.~\citeauthor{zhang2019deep}~\cite{zhang2019deep} addressed adversarial attacks as the drawback of applying deep learning solutions into mobile and wireless networking without covering security applications like network intrusion detection.~\citeauthor{martins2020adversarial}~\cite{martins2020adversarial} provided a limited review on some researches have applied the concepts of adversarial machine learning into malware and network intrusion detection contexts. In our previous paper we focused on the adversarial black-box attacks.   \label{relatedwork}
\section{Methodology} This study summarizes and investigates the state-of-the-art comprehensively to draw general conclusions and prelude further research. To elaborate this review, we followed the framework proposed by ~\citeauthor{kitchenham2004procedures}~\cite{kitchenham2004procedures} which involves three main stages: planning, conducting, and reporting. A review protocol was established for the planning stage that formulates the following elements: research questions, search strategy, study selection criteria, and data extraction strategy. 
\subsection{Research Questions}
This study investigates the threat of AEs towards ML-based NIDS and how this problem was addressed in the literature. Accordingly, this study tries to answer the following research questions:
\begin{itemize}
\item RQ1. How is AML applied in the NIDS domain? 
\item RQ2. What are the proposed countermeasures to mitigate the threat of AEs against ML-based NIDS in the literature?
\item RQ3. What are the criteria to assess the ability of adversarial attacks generation techniques in producing valid, realistic, and consistent adversarial flow?
\item RQ4. What is the feasibility of launching real-world adversarial attacks against ML-based NIDS?
\end{itemize}
The answers to these questions are provided in the discussion section~\ref{Research Questions}.
\subsection{Search Strategy}
We selected the following research keywords "adversarial machine learning" "OR" "Adversarial Attack" "AND" "network intrusion detection". The search was conducted in September 2021 on five selected scholarly databases: Scopus, Google Scholar, arXiv, WebofScience, and ResearchGate. It yielded into 130 papers. 
\subsection{Inclusion and Exclusion Criteria}
The retrieved research results were filtered according to formulated inclusion and exclusion criteria to exclude irrelevant papers. It is worth noting that all types of publications were included for filtering based on the predefined criteria, whether in workshops, symposiums, conferences, or journals. The exclusion criteria were to omit papers that are not written in English or accessible or do not address the AML attacks, particularly in the network intrusion detection domain. For instance, papers that apply AML in domains such as malware, medical images, 3D, time-series, and logs data were omitted. To screen the rest of the papers, the main inclusion criteria was to embrace a paper if it addresses the problem of AEs against at least one of the different DNNs architectures designed for NIDS. Moreover, the study needs to employ either supervised, semi-supervised, or unsupervised approaches as they are commonly implemented for NIDS, excluding the reinforcement learning-based ones. Additionally, the study needs to address the problem in one of the following networking environments: traditional networks, IoT, SDN, or WSN networks, excluding other types of networks such as industrial control systems (ICS) or Cyber-physical systems (CSPs). Accordingly, 61 papers in total were admitted as primary studies. The collected studies were analyzed and then categorized based on their application of AML into four groups in section~\ref{relatedwork}.

\subsection{Data Extraction}
In order to address research questions, a data extraction procedure was performed. We identified a list of characteristics to be extracted from each study in order to collect related information that can address our research questions and identify classification criteria for categorizing the literature. The description of each characteristic is provided in table~\ref{ClassificationCharacteristics}. 

\begin{table}[ht]
\centering
\setlength{\tymin}{70pt}

\begin{tabulary}{\columnwidth}{@{}|L|L|@{}}
\hline
Characteristic	&	Description	\\ \hline
Year	&	Year of the publication.	\\ \hline
Dataset	&	Dataset used for study evaluation.	\\ \hline
Model	&	ML algorithms used for study evaluation.	\\ \hline
Defense	&	Countermeasures utilized/proposed for mitigation AEs against ML-based NIDSs.	\\ \hline
Environment	&	Type of network traffic (i.e., Traditional, SDN, Wireless, or IoT).	\\ \hline
Setting	&	Assumed level of knowledge that the adversary has (i.e., white-box, black-box, or grey-box).	\\ \hline
Strategy	&	Phases of actions to lunch the attack (i.e., evasion, poisoning, oracle).	\\ \hline
Attack	&	Approach used for generating the AEs. 	\\ \hline
Results	&	Outcomes of the experimental evaluation in the form of measurements such as Accuracy, ROC, Evasion Rate, ..etc. 	\\ \hline
Purpose	&	Aim that study tries to achieve which can be to propose new techniques for generating adversarial network
flow, evaluate the resilience of ML-based NIDS models against AEs generated by generic adversarial attacks, or introduce countermeasures and to assess the effectiveness of some existing defensive mechanisms, solve the problems of data scarcity and data imbalance in network traffic.	\\ \hline

\end{tabulary}
\caption{Classification characteristics}
\label{ClassificationCharacteristics}

\end{table}
 \label{methodology}
\section{Adversarial Attacks Against ML-based NIDS} 

This section explores different studies that applied AML into the NIDS domain. The first category includes studies that propose new techniques for generating adversarial network flow that evades detection by ML-based NIDS models. The second group comprises studies that evaluate the resilience of a wide range of conventional and DL-based NIDS models toward AEs generated by several generic adversarial attacks. The third group of studies aims to introduce countermeasures and assess the effectiveness of some existing defensive mechanisms for securing ML-based NIDS against AML attacks. The last category applies AML to solve the problems of data scarcity and data imbalance that encounter building supervised and semi-supervised NIDS classifiers.  Due to the lack of real malicious traffic traces, AML approaches are utilized to generate synthetic malicious flow.
\subsection{\textbf{Generation of AEs to Attack ML-based NIDS Models}}
~\citeauthor{Li2018}~\cite{Li2018} proposed a technique that combines two types of adversarial strategies stealing model along with poisoning to attack ML-based NIDS in a black-box model. Firstly, an enhanced Synthetic Minority Oversampling Technique called Adaptive SOMTE (A-SMOTE) is implemented to generate new synthetic training data instances from an existing dataset with few labeled instances. This synthetic training dataset is then used to train a DNN substitute model. Finally, a proposed approach called Centre drifting Boundary Pattern (CBP) is utilized to craft AEs against the substitute model, which in turn, they have used to lunch a poisoning attack against the targeted NIDS model.  The attack's performance was evaluated using the NSL-KDD, Kyoto2006+, and WSN-DS datasets over NB, LR, SVM-sigmoid, SVM-RBF, and SVM-linear. The experimental results showed that the attack succeeded in reducing the accuracy rate on average to 82.30\% compared to  90.03\%          achieved by other techniques.

~\citeauthor{lin2018idsgan}~\cite{lin2018idsgan} introduced IDSGAN, a framework to generate adversarial examples that maintain a similar distribution of the original traffic but can deceive the model to make misclassifications. It is based on Wasserstein GAN that utilizes three components; a generator, discriminator, and black-box IDS. The generator crafts the adversarial malicious traffic by modifying certain features. The discriminator provides the feedback to assist the generator training and imitate the black-box attacks. The approach was evaluated using the NSL-KDD  dataset over SVM, NB, MLP, LR, DT, RF, and KNN classifiers. The experimental results showed that accuracy rates of all classifiers fall from 70\% on the original dataset to 1\% for all types of attacks. For making the attack more realistic, modifying the features was constrained to only the non-functional features that do not represent the functionality of the attack; such features can be changed or retained. On the other hand, the functional features representing the basic characteristics of the attack were kept unchanged, such as content-based features for the user-to-root attack or time-based features for the DoS attack. Even though the perturbation is limited to only the functionally and realistically changeable features, it was concluded that IDSGAN could effectively generate powerful AEs. Although preserving the network traffic's functional features was claimed, two functional features were altered which invalidate maintaining functional properties of adversarial traffic. Moreover, the training of GANs is currently unstable and suffers from model collapse and convergence failure.  

~\citeauthor{aiken2019investigating}~\cite{aiken2019investigating} explored the viability of adversarial attacks against an anomaly-based NIDS in an SDN environment. An adversarial testing tool named (Hydra) was introduced that creates adversarial evasion attacks for evading detection of TCP-SYN DDoS attacks by manipulating three features (i.e., bidirectional traffic, packet rate, and payload size). The produced AEs were assessed over Neptun, an ML-based detection platform for SDN that utilizes various ML classifiers and traffic features. The dataset consisted of benign traffic from the CICIDS2017 and malicious traffic from the DARPA SYN flood~\cite{gharaibeh2009}. Moreover, various classifiers were implemented for assessment, including FR, SV, LR, and KNN. The experimental results demonstrated that few perturbations for traffic features succeeded in crafting evasion attacks that can reduce TCP-SYN DDoS attacks' detection accuracy by Neptune classifiers from 100\% to 0\%. KNN proved to be the most resilient classifier. On the other hand, RF, LR, and SVM were vulnerable to the same perturbations; therefore, showing similar accuracy reductions. The drawback of this study is that the proposed tool can create adversarial examples for only one kind of saturation attack (i.e., TCP-SYN).

~\citeauthor{usama2019black}~\cite{usama2019black} introduced and validated a black-box adversarial attack design that compromises the integrity and availability of traffic classifiers and only demands querying the targeted model for labels. Mutual information (MI) is utilized to craft the adversarial perturbations, and substitute model training is used to lunch the adversarial attacks. The approach was evaluated over SVM and DNN classifiers using the UNB-CIC Tor dataset. The experimental results revealed that the accuracy rates dropped from 96.3\% to 2\% and from 96.4\% to 63.95\%, for DNN and SVM, respectively.

~\citeauthor{usama2019adversarial}~\cite{usama2019adversarial}  employed Mutual Information (MI) to extract the most discriminating features in both normal and DoS examples. Then, the distance between the most discriminating features of DoS examples is reduced using the constrained $l1$ norm minimization to craft the perturbations. The proposed attack was evaluated over SVM and DNN classifiers using the NSL-KDD dataset. The experimental results demonstrated that DNN's detection accuracy for DoS attacks dropped by 70.7\%.

~\citeauthor{peng2019adversarial}~\cite{peng2019adversarial}
analyzed DoS attacks' features to propose an enhanced boundary-based approach that generates adversarial DoS examples that can bypass ANN-based NIDS. The proposed method optimizes Mahalanobis distance by perturbing both continuous and discrete features of DoS instances. It works in a black-box setting by employing queries' outputs and optimization methods and takes DoS traffic characteristics into account. The method's performance was assessed over an ANN classifier using the KDDCup99 and CICIDS2017 datasets. The experimental results demonstrated that the proposed method could create adversarial DoS examples with limited queries and dropped the prediction confidence of the true class from 90\% to 49\%.

~\citeauthor{Yan2019}~\cite{Yan2019} proposed a DOS-WGAN architecture that utilizes Wasserstein GANs with gradient penalty to avoid detection by NIDS classifiers. DOS-WGAN automatically synthesizes DoS traffic characteristics based on the probability distribution of benign traffic. The proposed design was assessed over a CNN classifier using the KDDCup99 dataset. The evaluation showed that CNN's detection rate (i.e., true positive rate) dropped from 97.3\% to 47.6\%.  

~\citeauthor{Abusnaina2019a}~\cite{Abusnaina2019a} examined the suitability of applying the generic adversarial examples generating techniques to the domain of flow-based IDS in the SDN environments. The generic techniques are not suitable for crafting adversarial flow as the perturbed flow should be realistic and can be observed by actual packet flows, and flow-based features are highly interdependent, unlike in other domains (e.g., images) where features are independent. Therefore, FlowMerge was proposed, which addresses the challenges of generating adversarial flow that maintains the original flow traffic properties. FlowMegre changes the flow with a representative mask flow from a selected target class. Then, it combines the features of the original and masks flows using one of two methods; averaging or accumulating. This method is capable of fooling DL-Based IDS by offering both targeted and untargeted attacks. The performance of FlowMerge was compared to C\&W, ElasticNet, DeepFool, MIM, and PGD. The experiments over a CNN classifier demonstrated that the generic methods achieved a misclassification rate of 99.84\% for untargeted misclassification. On the other hand,  FlowMerge produced realistic adversarial flow with a success rate of 100\% for targeted misclassification. Moreover, adversarial training enhances the trained model against the AEs generated by the generic techniques but not by the FlowMerge. The shortcoming of the proposed method is that it was tested only over a CNN model, and it assumes that the adversary has complete knowledge about the classifier and traffic's features which is a persistent assumption.

~\citeauthor{apruzzese2019addressing}~\cite{apruzzese2019addressing} executed extensive experiments to evaluate the adversarial integrity attacks launched during the training-time and testing-time against three NIDS classifiers (i.e., RF, MLP, and K-NN) over the CTU-13 dataset. Under normal settings, the classifiers achieved recall measures between 0.93 and 0.97, with KNN the lowest performing and RF the highest. For the adversarial setup, a custom adversarial attack was implemented that adds random values within the interval of each one of three features which are exchanged\_bytes (number of bytes exchanged), duration (duration of the connections), and total\_packets (number of packets exchanged). The classifiers under the attack achieved lower recall measures between 0.31 for MLP neural network and 0.34 for RF. Moreover, the effectiveness of adversarial retraining and feature removal as countermeasures was evaluated. Using adversarial retraining, classifiers' performance enhanced and achieved recall measures between 0.49 for KNN and 0.60 for RF. Meanwhile, using feature removal, which eliminates perturbed features before training, the recall measure increased between 0.76 and 0.89 for MLP neural network and RF, respectively.

~\citeauthor{han2020practical}~\cite{han2020practical} presented an adversarial attack approach that can automatically mutate original traffic in black-box and grey-box settings while preserving traffic features functionality.  The proposed attack can be utilized to assess the robustness of different NIDSs using various ML/DL models and non-payload features. The approach utilizes GAN to produce the AEs, and PSO (i.e., an optimization procedure) to approximate these examples to the misclassification boundary. The experimental results over the Kitsune and CICIDS2017 datasets demonstrated that the proposed attack achieved evasion rates higher than 97\% in half of the cases. However, the adoption of GANs for AEs generation is computationally complex and time-consuming; therefore, such a technique is neither efficient nor practical for real-time attacks in contexts like IoT networks.

~\citeauthor{chen2020fooling}~\cite{chen2020fooling} introduced a feature generative framework named the Anti-
Intrusion Detection AutoEncoder (AIDAE), which learns the distribution of normal features and generates new random features; these features can bypass detection while preserving the distribution of normal ones without requiring feedback from targeted IDS during the training procedure. Moreover, it considers the correlation between the discrete and continuous features while generating the adversarial examples; therefore, it can be used for attacking the IDS in real scenarios. The framework utilizes an auto-encoder and a GAN to craft the adversarial examples. 
It was evaluated on NSL-KDD, UNSW-NB15, and CIC-IDS-2017 datasets over LR, KNN, DT, Adaboost, RF, CNN, and LSTM. Overall, detection rates dropped dramatically, achieving between 7.11 and 0.94.

~\citeauthor{alhajjar2020adversarial}~\cite{alhajjar2020adversarial} explored utilizing particle swarm optimization (PSO), genetic algorithm (GA), and GANs to generate AEs for network traffic. The performance of these generation techniques was assessed over SVM, DT, NB, KNN, RF, MLP, Gradient Boosting (GB), LR, Linear Discriminant Analysis (LDA), Quadratic Discriminant Analysis (QDA), and Bagging (BAG) using the NSL-KDD and UNSW-NB15 datasets. Moreover, their performance was compared to a baseline perturbation approach via Monte Carlo (MC) simulation, which generates random perturbations. The experiments demonstrated that the average evasion rates over all the classifiers on the NSL-KDD dataset were as follows: 96.19\%, 99.99\%, 99.95\%, 92.60\%, for MC, PSO, GA, GAN, respectively. 
For the UNSW-NB15 dataset, the average evasion rates over all the classifiers were 99.53\%, 98.06\%, 100\%, 99.61\%, for MC, PSO, GA, GAN, respectively. Overall, the best adversarial methods are GA, 
PSO, GAN, MC with an average evasion rate of 99.98\%, 99.02\%, 97.86\%, and 96.10\%, respectively.           


~\citeauthor{shu2020generative}~\cite{shu2020generative} introduced the Generative Adversarial Active Learning (Gen-AAL) algorithm that utilizes GANs with active learning to craft adversarial attacks against black-box ML-based NIDS. It requires a limited number of queries to the targeted model for labeled instances to train the GANs and does not demand a large training dataset. The proposed model was evaluated over a gradient boosted decision tree NIDS using the CICIDS2017 dataset. The experiments showed that the proposed model achieved a 98.86\% success rate in evading the NIDS classifier utilizing only 25 labeled instances during the training phase.


~\citeauthor{Teuffenbach2020}~\cite{Teuffenbach2020} presented a technique to generate AEs for NIDS that takes into account the domain constraints of network traffic features. The features are organized into groups, and each group is assigned a weight that indicates the feasibility of modifying them. The approach utilizes the C\&W attack to perform optimization in respect of perturbation constraints and feature weights. An attack's difficulty is expressed according to two kinds of constraints: feature space (feature budget) and the magnitude of feature alteration (perturbation budget). The approach creates a valid and realistic adversarial network flow as it restricts introducing perturbations to only accessible and independent features (modifying less than an average of 0.2). The method's performance was compared to FGSM and C\&W and evaluated over AE, DBN, and DNN classifiers using the NSL-KDD and CICIDS2017 datasets. Through the experiments, AE showed the highest robustness to AEs; however, it achieves the lowest accuracy rate in anomaly detection compared to DNN and DBN. On the other hand, DNN and DBN show more resistance to adversarial examples targeted to obscure DoS attacks than PortScan or Probe attacks.


~\citeauthor{chauhan2020polymorphic}~\cite{chauhan2020polymorphic} implemented polymorphic DDoS attacks using GANs in order to assess NIDS's ability at detecting adversarial examples and to enhance the training process for better resilience. The polymorphic DDoS attacks are generated by updating the DDoS attack profile features (i.e., number of features and swapping features), merging them with the previously created adversarial examples, and feeding them to the GAN model. The simulation results demonstrated that such polymorphic attacks, continuously change attacks profile, can avoid NIDS detection while maintaining a very low false-positive rate. Moreover, the defensive mechanisms that depend on incremental training remain vulnerable to new attacks. The approach was evaluated over the CICIDS2017 dataset using DT, LR, NB, and RF classifiers. The experiments demonstrated that detection rates dropped as low as 5.23\% and 3.89\% for changing the number of features and swapping features, respectively. The proposed model can be mainly utilized to generate a large volume of synthetic attacks that enrich the training process of NIDS models and to increase their detection for more recent attacks by being trained with those synthesized polymorphic attacks.


~\citeauthor{khamaiseh2020deceiving}~\cite{khamaiseh2020deceiving} investigated the resilience of some supervised and
unsupervised models against adversarial attacks in Software-defined Networking (SDN). An adversarial testing tool was proposed to craft adversarial evasion attacks that evade the detection of four saturation attacks (i.e., TCP-SYN, TCPSARFU, UDP, and ICMP). The traffic generator module is responsible for creating the evasion attacks by perturbing three traffic features (i.e., IPv4 source address change rate,  Ethernet source address change rate, and Packet rate). As a testing platform, several unsupervised and supervised classifiers (i.e., KNN, NB, SVM, IF, and ANN) were employed to assess generated evasion attacks' performance. The experiments were accomplished using a training dataset consisting of simulated and physical SDN traffic~\cite{khamaiseh2019detecting}. The experimental results showed that the detection performance for the four attacks was reduced by more than 90\% over all classifiers. For instance, recall of the KNN classifier decreased from 96\% to 8\%, and precision reduced by 91\%. With the least detection performance achieved by the IF classifier, recall reduced from 72\% to 5\%, and precision reduced from 76\% to 2\%.


~\citeauthor{qiu2020adversarial}~\cite{qiu2020adversarial} presented a realistic and feasible adversarial technique to attack DL-based NIDS in a black-box setting. The proposed technique utilizes the model extraction approach to replicate the targeted model for crafting AEs and demands only a small portion (10\%) of the original training dataset. Subsequently, it uses saliency maps to specify the most significant features that influence detection outcomes. Then, a gradient-based method (e.g., FGSM) is used to generate the adversarial examples by perturbing those crucial features. The method was implemented to attack Kitsune, an IDS for IoT networks. Two attack scenarios were demonstrated: An IoT botnet attack in which the attacker can alter malicious traffic to evade detection; and a video streaming application in which the attacker can alter benign traffic to force the detector to misclassify it as malicious and produce false-positive alarms. The traces of video injection and botnet malware attacks from the Kitsune dataset were used for the evaluation~\cite{mirsky2018kitsune}. The experimental results showed that the adversary could achieve an attack success rate higher than 95\%. It was concluded that the detection accuracy of NIDS could be significantly degraded with slight manipulation for the traffic.

~\citeauthor{Zhang2020}~\cite{Zhang2020} proposed a Brute-force Black-box Method to Attack Machine Learning-Based (BFAM) that overcomes the implementation complexity of GAN-based adversarial attacks and constructs the adversarial examples in an efficient and simple manner. The method works by tweaking the input vector in a controlled way and operates in a black-box manner without any internal information about the attacked model, making it a suitable technique to attack cybersecurity systems.  It is a gradient-free method and only needs the outputs from the targeted model and the confidence scores. Therefore, it can be used to assess the robustness of ML-based security solutions against adversarial examples in a black-box setting. 
The proposed approach's performance was assessed on various machine learning models, including NB, LR, RF, DT, and MLP, designed to detect Android malware, network intrusions, and host intrusions. The experiments showed that BFAM outperforms the adversarial attack method based on GAN (AAM-GAN), and RF classifiers are the most resilient to the AEs.

~\citeauthor{Yang2020}~\cite{Yang2020} proposed an optimization-based approach called universal adversarial sample generator (U-ASG) to launch white-box adversarial attacks on autoencoder-based semi-supervised network anomaly detection. The proposed method can launch attacks in a white-box manner where the adversary has detailed knowledge, including the datasets, parameters, and threshold of the trained autoencoder model. The attacker aims to do small perturbations to the changeable features (e.g., some features in network traffic are unchangeable like service, protocol, etc.) of a detected anomalous instance to fool the model to misclassify it as a normal. The approach's effectiveness was evaluated on deterministic autoencoder (DAE) and variational autoencoder (VAE) network anomaly detection models using the KDDCUP99 dataset. To measure the performance of the generated AEs, the authors used a metric named anomalous-to-normal rate (ANR), which refers to the proportion of the AEs that are wrongly classified as normal. The experiments revealed that the generated AEs were more effective on DAE than those on VAE. The ANR of U-ASG is higher than AT-VAE ~\cite{gondim2018adversarial} by around 20\%.


~\citeauthor{kuppa2019black}~\cite{kuppa2019black} proposed a more realistic black-box attack against various NIDS classifiers in query-limited and decision-based settings. The attack learns and approximates the distribution of benign and anomalies instances to avoid NIDS with a high success rate. It utilizes Manifold Approximation Algorithm for query reduction and spherial local subspaces for generating the adversarial examples. Manifold Approximation Algorithm is applied to the targeted model's collected responses at the attacker end for query reduction and identifying the anomaly detectors' thresholds. Then, spherial local subspaces is used to produce the AEs. This approach is suitable for attack anomaly detectors whose decisions are threshold-based, and in case boundaries between anomaly and normal classes are not well defined. The proposed attack method is evaluated over the CICIDS2018 using Autoencoder, Deep Autoencoding Gaussian Mixture Model (DAGMM)~\cite{zong2018deep}, AnoGAN~\cite{schlegl2017unsupervised}, Adversarially Learned Anomaly Detection (ALAD)~\cite{zenati2018adversarially}, Deep Support Vector Data Description (DSVDD)~\cite{ruff2018deep}, One-Class Support Vector Machines (OC-SVM)~\cite{scholkopf2000support}, and Isolation Forests (IF)~\cite{liu2008isolation}. Overall, the experimental results showed that the attack achieved over 70\% success rate on all classifiers.


~\citeauthor{wang2020c}~\cite{wang2020c} addressed the unsuitability of applying AEs generation techniques of image processing to the NIDS domain, which have been used by most of the studies. A Constraint-Iteration Fast Gradient Sign Method (CIFGSM) was implemented to adapt to network traffic complexity, different types of features, and correlations among features. The performance of CIFGSM was compared to the IFGSM over the NSL-KDD dataset and using DT, CNN, and MLP classifiers. The experiments showed that CIFSM outperformed IFGSM in terms of classification accuracy, feature type matching, Euclidean metric, and rank of the matrices. The accuracy rates for DT, CNN, and MLP classifiers under the CIFGSM attack dropped to 0.25, 0.68, and 0.73, respectively.
~\citeauthor{cheng2021packet}~\cite{cheng2021packet} proposed Attack-GAN that crafts adversarial traffic at the packet level that maintains the domain constraints. The Attack-GAN utilizes Sequence Generative Adversarial Nets (SeqGAN) with policy gradient in which the generation of adversarial packets is constructed as a sequential decision-making process. The proposed technique was evaluated using the CTU-13 dataset over MLP, RF, DT, and SVM in a black-box setup. Although Attack-GAN can generate adversarial examples that evade detection, it can only generate fixed-length packets.


~\citeauthor{sharon2021tantra}~\cite{sharon2021tantra} presented TANTRA, a timing-based adversarial network traffic reshaping attack that reshapes malicious traffic using timestamp attributes in order to evade detection without affecting the packet’s content. The approach was assessed using the Kitsune and CIC-IDS2017 datasets over the KitNET, an advanced NIDS, Autoencoder, Isolation Forest, and achieved a success rate of 99.99\%. Furthermore, a mitigation solution was proposed that involves training the model with both benign and reshaped traffic and has shown promising results.  


~\citeauthor{guo2021black}~\cite{guo2021black} introduced a black-box setup for generating adversarial network traffic. Firstly, a substitute model is trained with a similar decision boundary to the target model. Then, the BIM is extended to generate adversarial traffic with the parameters and the structure of the substitute model. The method was evaluated using the KDD99 and CSE-CIC-IDS2018 over CNN, SVM, KNN, MLP, and Residual Network (ResNet).


~\citeauthor{gomez2021crafting}~\cite{gomez2021crafting} proposed Selective and Iterative Gradient Sign Method (SIGSM) that overcomes the limitations of existing adversarial attack generation methods by controlling the features to be manipulated. 
The method was designed for industrial control system environments (ICS), and its performance was evaluated using the Electra dataset, collected from an Electra Traction Substation. The experiments showed that the resulting adversarial examples reach their destination as the network's intermediate devices can understand them, and therefore they can lead to a successful real-world attack in industrial scenarios. The 
SIGSM was compared with FGSM and BIM regarding the number of perturbed features and execution time. The evaluation demonstrated that FGSM and BIM transformed all the features such as MAC and IP addresses, while SIGSM produced an identical feature vector to the original except payload data. In terms of execution time, FGSM was the fastest, and SIGSM 
took less time than BIM because it applies the clipping operation on selected features rather than all the features as in the BIM attack.  

~\citeauthor{anthi2021hardening}~\cite{anthi2021hardening} utilized InfoGain Ratio, a feature importance ranking method, to discriminate between benign and malicious traffic. Accordingly,  the top-ranked features were selected to be perturbed. The performance of J48 DT, RF, Bayesian Network, and SVM was evaluated under adversarial examples generated by manipulating all of the important features together and one at a time. However, the proposed approach is only applicable for generating adversarial examples for DoS attacks and targeting only supervised-based ML NIDS.

~\citeauthor{shiehdetection}~\cite{shiehdetection} utilized Wasserstein Generative Adversarial Networks with Gradient Penalty (GP-WGAN) to craft synthesized DDoS adversarial traffic. The experiments revealed that the produced adversarial traffic could penetrate MLP,  RF, and K-NN without being detected. As a defensive mechanism, the study proposed an adversarial GAN intrusion detection (AG-IDS) with dual discriminators, where the addition is to detect adversarial DDoS flow. The adversarial detection rate of AG-IDS over adversarial traffic reached up to 87.39\%, compared to 0\% for the original GAN and ANN models.

\subsection{\textbf{Evaluation of ML-based NIDS Model Resilience to AEs}}
~\citeauthor{rigaki2017adversarial}~\cite{rigaki2017adversarial} studied the suitability of FGSM and JSMA approaches to generate effective targeted adversarial examples that evade ML-based NIDSs. In a grey-box setting, the AEs were generated against an MLP substitute classifier and then were transferred to attack other models. The study evaluated the robustness of some well-known NIDS classifiers (i.e., DT based on the CART algorithm, SVM with a linear kernel, RF, and the Majority Voting ensemble method) towards the generated AEs. Using the JSMA method, the experimental results showed that all the classifiers were affected and showed an accuracy degradation from 73\% to 45\% in the best case, with linear SVM the most affected with a drop of 27\% in accuracy, and RF is the most resilient with a drop of 18\% in the accuracy rate and 6\% in F1 score and AUC.  It was concluded that FGSM is an unsuitable approach to evade ML-based NIDS as it altered 100\% of the traffic features; on the other hand, JSMA is a more realistic attack as it modified on average 6\% of the features. The major limitations of this study are that knowledge about the target classifier's features is assumed, and the attack approach generates the feature vectors, not the  AEs themselves.
~\citeauthor{warzynski2018intrusion}~\cite{warzynski2018intrusion} evaluated the resilience of an FNN binary classifier against the FGSM attack over the NSL-KDD dataset. The experiments revealed that the classifier's performance was degraded significantly with a False-Negative rate equal to zero, indicating that all adversarial instances have been misclassified as benign.


~\citeauthor{Wang2018}~\cite{Wang2018} utilized state-of-the-art white-box adversarial attacks (i.e., FGSM, JSMA, DeepFool, and C\&W) to evaluate the robustness of an MLP classifier. The classifier achieved an AUC of 0.94 over the clean samples of the NSL-KDD dataset. The four attacks effectively reduced the classifier's AUC in the adversarial setting, with FGSM being the most effective and C\&W the least. Overall, the study showed that the adversarial examples could degrade DoS attack detection accuracy from 93\% to 24\%. The values of AUC are suppressed to 0.44, 0.5, and 0.80 under FGSM, JSMA, and C\&W, respectively. Moreover, the features that contribute to generating the AEs among the different adversarial attacks are identified. The attacks primarily use seven features, which therefore demand better protection to prevent exploitation by adversaries. From an applicability perspective, the attacks that alter a limited number of features are more attractive to attackers. The JSMA attack was found to be the most realistic attack since it perturbs a small range of features. The author also discussed how attackers could change each of the seven features manually in real scenarios.

~\citeauthor{Yang2019}~\cite{Yang2019} used three black-box attacks based on a training substitute model, ZOO, and Wasserstein Generative Adversarial Network (WGAN) to demonstrate the attacker's ability to generate effective AEs without any internal information about the targeted model. For the first attack, a substitute model, similar to the victim model, is trained with the same training dataset, and the C\&W method is used to generate the AEs, due to its efficiency. Because of adversarial examples' transferability, the generated AEs for the substitute model are used to attack the targeted model. In the second attack, the ZOO method ~\cite{chen2017ZOO} queries the detector to estimate the gradient in order to generate the AEs.  For the last attack, a WGAN~\cite{goodfellow2014generative} is utilized to craft the AEs. Overall, an FNN model's performance over the NSL-KDDD dataset under the three attacks was significantly reduced, which indicates the possibility of launching highly damaging attacks  without any internal information about the detection model. The attacks achieved F1 score degradation of 70\%, 62\%, and 24\% for ZOO, WGAN, and C\&W, respectively. It was concluded that the substitute model attack is the least effective, and the ZOO attack the most. However, it is computationally expensive as it requires querying the detector to update the gradients, which may reduce its effectiveness in real scenarios where the NIDS limits the number of queries. Additionally, the GAN showed a good performance; however, training GANs was unstable and suffered from convergence failure and model collapse. The study's drawback is that it handles the network traffic dataset arbitrarily and does not constrain the perturbations for crafting the AEs.  

~\citeauthor{clements2019rallying}~\cite{clements2019rallying} investigated the robustness of a recently introduced light-weight DL-NIDS for IoT networks, Kitsune. The resilience of Kitsune was evaluated against FGSM, JSMA, C\&W, and ENM attacks using the Mirai dataset. The experimental results showed that the integrity attacks using these algorithms achieved success rates of 100\%. Additionally, C\&W and ENM achieved success rates of 100\% in the availability attacks. On the other hand, FGSM and JSMA performed worse, with a success rate of 4\% and 0\%, respectively. It was concluded that by altering as few as 1.38 of the traffic features on average, the attacker could craft adversarial examples capable of bypassing the DL-based NIDS.


~\citeauthor{Peng2019}~\cite{Peng2019} proposed Evaluating Network Intrusion Detection System (ENIDS) 
framework to assess the robustness of ML-based NIDS against AEs. Four adversarial attacks PGD, MI-FGSM, L-BFGS, and SPSA were used to generate the AEs. The resilience of several well-known classifiers (i.e., DNN, SVM, RF, and LR) towards adversarial attacks was assessed over the NSL-KDD dataset. The models showed different resilience to AEs, with DNN achieving the lowest ROC metric of 0.37 and the MI-FGSM attack achieving the best performance.  


~\citeauthor{Ibitoye2019}~\cite{Ibitoye2019} compared the resilience of SNN to FNN against AEs generated by FGSM, BIM, and PGD attacks in IoT networks. The experimental results over the BoT-IoT dataset revealed that FNN outperforms SNN in terms of accuracy metrics. However, SNN showed more robustness to AEs. Moreover, the study investigated the effects of feature normalization on accuracy rate and resilience to AEs of the two models. It was found that both models' accuracy detection rates on the non-normalized datasets have been lowered significantly; however, their robustness to the AEs was enhanced.


~\citeauthor{jeong2019adversarial}~\cite{jeong2019adversarial} evaluated the performance of AE and CNN over the NSL-KDD dataset against adversarial examples of the FGSM and JSMA attacks. The experiments demonstrated that the performance of both models degraded significantly. The AE's accuracy rate was reduced on average by 48.16\% and 49.48\% for the FGSM and JSMA attacks, respectively. Meanwhile, the CNN model showed an accuracy rate reduction on average by 44.33\% and 55.15\% for the FGSM and JSMA attacks, respectively.


~\citeauthor{Huang2019}~\cite{Huang2019} tested the performance of three port-scan attack detecting models MLP, CNN, and LSTM, in an SDN environment under three adversarial attacks; FGSM, JSMA, and JSMA-RE (JSMA reverse). The built models detect port-scanning attacks based on features extracted from the OpenFlow protocol messages (PACKET\_IN) and controller's flow statistics (STATS). Overall, the JSMA attack results in the most significant drop in the performance of the three models, ranging from 14\% to 42\%. Meanwhile, FGSM overall does not perform well; however, it caused a significant drop in LSTM's accuracy of more than 50\%. The JSMA-RE attack achieved a 35\% drop in the MLP model's accuracy with no significant effect on either the CNN or LSTM models' accuracy rates.

~\citeauthor{martins2019analyzing}~\cite{martins2019analyzing} studied the resilience of six well-known NIDS classifiers (i.e., DT, RF, SVM, NB, NN, and DAE) against AEs that are generated by FGSM, JSMA, Deepfool, and C\&W approaches. The evaluation was carried over the DoS traces only present in the NSL-KDD and CICIDS2017 datasets. The study showed that all attacks could degrade the classifiers' performance and reduce average AUC by 40\% on CICIDS2017 and 13\% on NSL-KDD. Achieving a significant AUC decrease over the CICIDS2017 dataset was due to the presence of a large portion of modifiable features that can be exploited compared to those in the NSL-KDD dataset. Notably, DAE was the most robust classifier showing a decline in AUC by only 1\% for each of the two datasets. Moreover, the performance of all models was enhanced by using adversarial training as a defensive mechanism. Distinctly, RF was the best performing model on the two datasets as it suffered only 0.1\% reduction of AUC between the original and manipulated versions of each dataset.

~\citeauthor{piplai2020nattack}~\cite{piplai2020nattack} assessed a GAN classifiers' resilience to the AEs. The GAN was used to generate AEs during the training stage, aiming to make it more resistant to the AEs. Then, the FGSM approach was utilized to launch adversarial attacks on the model. The model was evaluated over the IEEE BigData 2019 cup: Suspicious network event recognition ~\cite{janusz2019ieee}. The experiments showed that the GAN's performance degraded under the adversarial attack, and the lowest achieved attack success rate was 41\%, which indicates that adversarial training cannot be an effective defensive mechanism.

~\citeauthor{Sriram2020}~\cite{Sriram2020} analyzed the performance of several ML and DL models that are widely implemented for network intrusion detection using the NSDLKDD dataset in adversarial and non-adversarial environments. The AEs were generated by using two of the most common attack techniques; (e.i., FGSM and JSMA). In this comparison study, the built classifiers are LSTM, CNN, DNN, SVM, NB, KNN, LR, DT, RF, AB,  and SMR. For the FGSM attack, the top three most affected classifiers are DNN, LSTM, and DT, with performance reductions of 78\%, 76\%, and 67\%, respectively. The least affected classifiers by the FGSM are RBF-SVM, LR, and LSVM, with performance reductions of 2\%, 2\%, and 4\%, respectively. Meanwhile, for the JSMA attack, the top three most affected classifiers are CNN, DNN, and DT, with performance reductions of 87\%, 86\%, and 83\%, respectively. On the other hand, the least affected classifiers by the JSMA are NB, RBF-SVM, and LSVM, with performance reductions of 2\%, 4\%, and 30\%, respectively. Overall, the FGSM works better than JSMA in the case of LSTM and NB, but JSMA outperforms the FGSM in the other cases. Moreover, RBF-SVM, LSVM, KNN, and NB are the most resilient classifiers against both adversarial attacks.

~\citeauthor{zhong2020adversarial}~\cite{zhong2020adversarial} proposed MACGAN (maintain attack features based on the generative adversarial networks) to assess the resilience of ML-based NIDS and to enhance their stability. The framework consists of two major components; the first is to analyze attack features manually, and the second is the GAN which is utilized to evade the detection models.  The traffic features are divided into perturbable and non-perturbable to maintain original traffic characteristics, and the generator modifies only perturbable ones. The framework was tested on the recent CICIDS2017 and Kitsune2018 data sets over various NIDS classifiers, including Kitsune Anomaly Detection Algorithm, IF, GMM, SVM,  SAE, and RBM. The experiments showed that the Kitsune algorithm achieved the highest TPR, almost up to 0.998, before being attacked. However, after being attacked, its TPR  nearly decreased to zero. The other algorithms showed the same trend.

~\citeauthor{pacheco2021adversarial}~\cite{pacheco2021adversarial} evaluated the performance of DT, SVM, and RF under JSMA, FGSM, and CW attacks. The evaluation study was performed over the BoT-IoT and UNSW-NB15 datasets. The attacks demonstrated varied results on the two datasets, with JSMA the least effective on both datasets, CW the most effective on the UNSW-NB15, and FGSM the most effective on the BoT-IoT dataset. Furthermore, the RF was the most robust classifier, and SVM was the least on both datasets.

~\citeauthor{debicha2021adversarial}~\cite{debicha2021adversarial} examined the impact of FGSM, BIM, and PGD attacks on a DNN model for NIDS. The experimental results over the NSL-KDD dataset showed the FGSM decreased the model's accuracy from 99.61\% to 14.13\%, while PGD and BIM decreased it further to 8.85\%. Furthermore, the study evaluated the effectiveness of adversarial training (AT) as a defensive method. Although AT can improve a model's resilience to adversarial examples, it can slightly decrease the accuracy rate on the clean (unattacked) data.  

~\citeauthor{maarouf2021evaluating}~\cite{maarouf2021evaluating} assessed the resilience of encrypted traffic classification models against adversarial evasion attacks, namely ZOO, PGD, and DeepFool. The performance of C4.5, KNN, ANN, CNN, and RNN classifiers were compared in adversarial-free and adversarial attack environments using two benchmark datasets for encrypted traffic; SCX VPN-NonVPN~\cite{draper2016characterization} and NIMS~\cite{alshammari2007flow}. Overall, DL-based models demonstrated better classification in the adversarial-free setup compared to conventional ML. However, in the presence of adversarial attacks, the robustness of DL/ML depends on the attack's type. Furthermore, DeepFool was the most effective attack on both types of models.

~\citeauthor{fu2021robust}~\cite{fu2021robust} tested the robustness of LSTM, CNN, and Gated (GRU) against AEs generated by FGSM attack. The resilience of the models was assessed over the CICIDS2018 dataset using three training setups; training with normal examples, adversarial examples, and third pretraining with normal examples and training with adversarial examples. Their evaluation showed that adversarial training enhanced the resilience of the models, with LSTM being the most improved. However, AT reduces the accuracy of the models on the normal examples. While AT makes models’ decision boundaries more adaptable to AEs, it fragilizes the decision boundary of normal examples, which relatively undermines these examples' judgment.

~\citeauthor{nugraha2021detecting}~\cite{nugraha2021detecting} investigated the performance of MLP and CNN-LSTM using two adversarial datasets; one derived from an emulated SDN network, and the other synthetically produced using Tabular GAN. Furthermore, the study evaluated the effectiveness of adversarial training in three setups. The experimental evaluation showed that the more adversary examples are included in the training dataset, the more robust the MLP became.



\subsection{\textbf{Defending ML-based NIDS Models' Against AEs}}

~\citeauthor{AbouKhamis2019}~\cite{AbouKhamis2019} investigated the resilience of DL-based NIDS towards AEs. The min-max method was utilized for training a DNN model against AE over the UNSW-NB15 dataset. The max approach was utilized to generate AEs that maximize loss. On the other hand, the min approach was utilized as a defensive mechanism that minimizes the loss of injected AEs during adversarial training. The AEs were generated using Randomized Rounding Approach (rFGSM\textsuperscript{S}), Deterministic Approach (dFGSM\textsuperscript{S}), Multi-Step Bit Gradient Ascent (BGA\textsuperscript{S}), and Bit Coordinate Ascent (BCA\textsuperscript{S}).  For each approach, a model was trained with its produced AEs and benign examples to enhance the model's robustness against AEs. Additionally, a natural model is trained with a clean dataset. For the testing, the five built models were attacked by four sets of AEs for each technique. Overall, the (dFGSM\textsuperscript{S}) model (i.e., trained with AEs generated by dFGSM\textsuperscript{S}) achieved the lowest evasion rates across all the adversarial attack techniques. Meanwhile, the (BGA\textsuperscript{S}) outperforms all attack methods across the five built models. In a second setting of the experiment, Principle Component Analysis (PCA)  was applied for feature reduction in data pre-processing. Predominantly, utilizing PCA boosted the robustness of trained models against AEs. The evasion rates were approximately three times lower compared to the first experiment. Moreover, the trained models demonstrated equal resilience against all types of AEs. It was concluded that a further exploration for dimensionality reduction in DL could enhance the resilience of DL-based NIDS against evasion attacks.

~\citeauthor{usama2019generative}~\cite{usama2019generative} utilized GANs to generate AE while maintaining functional behavior and altering only non-functional properties of network traffic. Additionally, GAN-based training was proposed as a defensive approach in which the network is injected with AEs during the training phase to enhance its robustness. The performance of GAN-based AEs was evaluated over DNN, LE, SVM, KNN, NB, RF, DT, and GB using the KDDCup99 dataset. The experiments demonstrated that the highest accuracy was 65.38\% for GB, and the lowest was 43.44\% for SVM. After GAN-based training, the classifiers' performance was notably enhanced, achieving accuracy rates between 86.64 and 79.31 for LR and KNN, respectively.


~\citeauthor{benzaid2020robust}~\cite{benzaid2020robust} proposed an application-layer DDoS self-protection framework that is robust towards adversarial examples. The framework leverages DL and SDN enablers to empower fully autonomous mitigation and detection for the application-layer DDoS attacks. The proposed framework demonstrated effective performance in terms of system load and server response time. The detector is built using MLP and trained with legitimate and DDoS traffic traces from the CICIDS2017 dataset. Adversarial training was employed as a defensive approach where the model is trained with AEs generated by the FGSM technique.

~\citeauthor{hashemi2020enhancing}~\cite{hashemi2020enhancing} presented Reconstruction from Partial Observation (RePO) as a mechanism for enhancing the robustness of an unsupervised denoising autoencoder anomaly detector for packet and flow-based NIDSs towards AEs. The proposed method was compared to Kitsune~\cite{mirsky2018kitsune}, BiGAN~\cite{zenati2018efficient}, DAGMM~\cite{zong2018deep} as baselines, and the adversarial attacks were crafted by an approach proposed in~\cite{hashemi2019towards}. The assessment results over the CICIDS2017 dataset showed that the proposed mechanism improved malicious traffic detection by 29\% and 45\% in the normal and adversarial settings, respectively.

~\citeauthor{zhang2020tiki}~\cite{zhang2020tiki} presented (TIKI-TAKA), a framework that assesses and increases the resistance of DL-based NIDS against adversarial evasion attacks. In the evaluation, three DL-based NIDS classifiers (i.e., MLP, CNN, and LSTM) were attacked using five different adversarial methods (i.e., NES, Boundary, HopSkipJumpAttack, Pointwise, and Opt-Attack). The utilized techniques generate AEs that maintain the domain constraints and the semantics of a realistic network traffic flow. The evaluation revealed that attackers could bypass NIDS with an attack success rate that reaches up to 35.7\% under realistic conditions by amending the time-based features only. Moreover, three defending approaches were introduced that include: ensembling adversarial training, query detection, and model voting ensembling, which can be used individually or jointly to decrease the success rates of evasion attacks.

~\citeauthor{Qureshi2020}~\cite{Qureshi2020} introduced a Random  Neural Network-based Adversarial intrusion detection (RNN-ADV) that is trained with Artificial Bee Colony (ABC) Algorithm to detect adversarial attacks. The proposed model's performance against JSMA attacks was compared to DNNs over the NSL-KDD dataset.  In the adversarial environment, RNN-ADV reported an F1 score of 52.60\% for the normal class and 44.65\% for the DoS class.  Meanwhile, DNN reported F1 scores of 35.69\% and 25.89\%, respectively. Overall, RNN-ADV performed better and classified AEs with higher accuracy and precision.  

~\citeauthor{Apruzzese2020}~\cite{Apruzzese2020} introduced (AppCon) an integratable approach to the existing real-world defensive mechanisms that protect NIDS against adversarial evasion attacks while retaining effective performance in their absence. The proposed solution restricts the range of the samples that adversaries can craft to avoid the detector and leverages ensemble learning as a defensive strategy.  The AEs were generated by modifying combinations of malicious flow-based features incrementally in small amounts~\cite{apruzzese2018evading}. The solution's effectiveness was assessed over well-known NIDS classifiers, including DT, RF, Adaboost, MLP, and Wide and Deep (WnD) using the CTU-13 dataset. The extensive experiments showed that AppCon could reduce the success rate of evasion attacks against botnet detectors by almost 50\% while preserving an efficient detection performance in the non-adversarial setting.

~\citeauthor{pawlicki2020defending}~\cite{pawlicki2020defending} proposed an adversarial detector that utilizes neural activations to detect adversarial attacks at the inference phase. The neural activations of an ANN trained with a part of the CICIDS2017 dataset were collected at the testing. These collected neural activations were used to train and test various classifiers to detect the AE (i.e., Adaboost, ANN, RF, and SVM). The experiments demonstrated a recall measure of 0.99 for the adversarial evasion attacks by RF and KNN. The evasion attacks were crafted using FGSM, BIM, C\&W, and PGD algorithms.

~\citeauthor{peng2020detecting}~\cite{peng2020detecting} presented an adversarial sample detector (ASD) that utilizes bidirectional generative adversarial network (BiGAN) for defending NIDS against adversarial attacks. The BiGAN comprises a generator, discriminator, and encoder.  During training, the generator learns the distribution of benign examples. After training, the ASD module computes reconstruction errors, and the discriminator matches the errors of samples. The ASD module detects adversarial samples and eliminates them before they are fed to the NIDS. The experimental results revealed that DNN's accuracy rate was reduced by 60.46\%, 28.23\%, and 46.5\% under FGSM, PGD, and MI-FGSM, respectively.  With ASD, the accuracy rate was improved by 26.46\% against the PGD attack and by 11.85\% against FGSM. However, the impact of ASD on IM-FGSM examples was not clear.

~\citeauthor{abou2020evaluation}~\cite{abou2020evaluation} evaluated the min-max formulation~\cite{madry2017towards}, which augments crafted inputs during model training, as a defensive mechanism. The AEs were generated using five white-box attacks (i.e., FGSM, CW, BIM, PGD, and DeepFool). The min-max formulation's effectiveness was assessed on ANN, CNN, and Recurrent Neural Network classifiers over the UNSW-NB15 and NSD-KDD datasets.  The experimental results revealed that the accuracy rates of the classifiers on average were enhanced from 39\% to 92\%.

~\citeauthor{wang2021def}~\cite{wang2021def} proposed Def-IDS, an ensemble defense mechanism that protects NIDS from both known and unknown adversarial attacks.  The proposed ensemble retraining mechanism incorporates two modules: a multi-class generative adversarial network (MGAN) that crafts mimic examples for the intrusion instances using a single GAN model; the second module is multi-source adversarial retraining (MAT) that integrates adversarial examples generated by various attacks (FGSM, DeepFool, JSMA, and BIM). The effectiveness of this approach was assessed over a DNN classifier using the CSE-CIC-IDS2018 dataset, compared with other defensive mechanisms, and showed its ability to detect different adversarial attacks (namely, FGSM, DeepFool, JSMA, and BIM) with better accuracy rates.  

~\citeauthor{novaes2021adversarial}~\cite{novaes2021adversarial} employed GAN as DDoS detection in an SDN environment. Due to its nature, GAN can generate adversarial traffic and apply adversarial training as a defensive mechanism. GAN outperformed MLP, CNN, and LSTM models over the CICDDoS 2019 dataset and emulated real SDN traffic.

~\citeauthor{mccarthy2021feature}~\cite{mccarthy2021feature} assessed feature selection for model training in improving the models' resilience against adversarial examples generated by FGSM attack. The experiments utilized the traces of normal and DDoS from the CICIDS2017. The study analyzed the discriminator features between the original and adversarial traffic generated by FGSM using dimensionality reduction methods: PCA, t-SNE, UMAP, and Parallel coordinates. 
Accordingly, Recursive Feature Elimination (RFE) was used to remove the features with the largest absolute difference under the FGSM attack, and the detection model was retrained without them. The evaluation showed that the model achieves the highest accuracy rate when most features are used. However, the accuracy rate rarely exceeds  60\% under the attack. Furthermore, the size of the required perturbation for a successful attack tends to decrease when more features are incorporated. The addition of features expands surface attack and allows more subtle perturbations. Using feature removal, the model achieved an accuracy rate of 86.86\% over the FGSM’s adversarial examples with no drop in accuracy over the unperturbed examples.

~\citeauthor{ganesan2021mitigating}~\cite{ganesan2021mitigating} utilized feature selection (namely, RFR) to remove the features that are targeted by the adversarial evasion attack to confuse the detection model. Multiple feature sets of reduced size were used to identify an ensemble of multiple ML models (i.e., DNN, SVM, LR, and RF) that is more resilient to evasion attacks. The experiments were performed using KDDCup99, CICIDS, and DARPA datasets, and the Hydra tool was used to generate the evasion dataset.  The experimental evaluation demonstrated that the ensemble classifier could effectively detect several evasion attacks that cannot be detected using a single classifier with the complete set of features.   

~\citeauthor{debicha12detect}~\cite{debicha12detect} studied the transferability of black-box adversarial attacks over various ML-based NIDSs that utilize different ML techniques in black-box settings. FGSM and PGD were used to generate adversarial examples to evade a DNN model. Then, the transferability of these examples among popular ML algorithms (i.e., SVM, DT, LR, RF, and LDA) was examined. The experiments revealed that although the DNN model was the most deteriorated, the other models suffered some deterioration but with different degrees due to their composition of differentiable elements. Moreover, the study showed that using an ensemble of detection models was more robust against the transferable attacks compared to a single model. Additionally, the effectiveness of the Detect \& Reject approach was examined for mitigating the impact of transferability of adversarial attacks among ML-based detection classifiers.

\subsection{\textbf{Synthesize Malicious Network Traffic}}
~\citeauthor{zhang2019deep1}~\cite{zhang2019deep1} utilized deep adversarial learning and statistical learning to solve two significant problems that face supervised and semi-supervised NIDS classifiers, which are data scarcity and data imbalance. 
DL-based NIDSs demand large training datasets to optimize the weights. A data augmentation (DA) module was proposed to address the aforementioned challenges that can be part of a NIDS framework. The module includes a Poisson-Gamma Joint Probabilistic Generative model to estimate feature distributions of network intrusions data, a Monte Carlo method to generate synthesized data, and finally, a deep GAN to augmented the synthesized data. The framework was evaluated over the KDDCUP99 dataset using DNN, SVM, and LR. The assessment showed that the DA improved the classifier's performance and outperformed other approaches in terms of F1-score especially and other metrics.

Table~\ref {table:Summary of surveyed studies} shows a summary of the studies that have been covered in this survey, and a brief overview of utilized datasets is presented in table \ref{table:Overview of used datasets in the literature}.
 \label{literaturereview}
\section{Discussion} \begin{table*}[t]
\centering
\setlength{\tymin}{70pt} 

\begin{tabulary}{\textwidth}{@{}|L|L|L|L|L|@{}}
\hline
        \textbf{Ref.} & \textbf{Dataset} & \textbf{Network} & \textbf{Year} & \textbf{Attacks Categories} \\ \hline
~\cite{KDDCup99}	&	KDD CUP 99	&	Traditional 	&	1999	&	DoS, Probe, User 2 Root and Remote to User	\\ \hline
~\cite{universityofnewbrunswickest.1785}	&	NSL-KDD	&	Traditional 	&	2009	&	DoS, Probe, User 2 Root and Remote to User	\\ \hline
~\cite{gharaibeh2009}	&	DARPA	&	Traditional 	&	2009	&	DDoS, Malware, Spambots, Scans, Phishing	\\ \hline
~\cite{garcia2014empirical}	&	CTU-13	&	Traditional 	&	2011	&	Botnet 	\\ \hline
~\cite{KyotoDataset}	&	Kyoto	&	Traditional 	&	2015	&	Botnet 	\\ \hline
~\cite{moustafa2015unsw}	&	UNSW-NB15	&	Traditional 	&	2015	&	Backdoors, Fuzzers, DoS, Generic, Shell code, Reconnaissance, Worms, Exploits, Analysis	\\ \hline
~\cite{almomani2016wsn}	&	WSN-DS	&	Wireless 	&	2016	&	Greyhole, Blackhole, Scheduling, Flooding.	\\ \hline
~\cite{niyaz2016deep}	&	SDN traffic	&	SDN 	&	2016	&	DDoS	\\ \hline
~\cite{sharafaldin2018toward}	&	CICIDS2017	&	Traditional 	&	2017	&	DoS, DDoS, SSH-Patator, Web, PortScan, FTP-Patator, Bot.	\\ \hline
~\cite{antonakakis2017understanding}	&	Mirai	&	IoT	&	2017	&	Botnet 	\\ \hline
~\cite{sharafaldin2018toward}	&	CICIDS2018	&	Traditional 	&	2018	&	Bruteforce Web, DoS, DDoS, Botnet, Inﬁlteration.	\\ \hline
~\cite{koroniotis2019towards}	&	BoT-IoT	&	IoT	&	2018	&	DDoS, DoS, OS Service Scan, Keylogging, Data exfiltration	\\ \hline
~\cite{mirsky2018kitsune}	&	Kitsune	&	IoT	&	2018	&	Recon, Man in the Middle, DoS, Botnet Malware	\\ \hline
~\cite{janusz2019ieee}	&	IEEE BigData cup	&	Traditional 	&	2019	&	N/A	\\ \hline

    \end{tabulary}
    \caption{Overview of used network intrusions datasets in the literature}
\label{table:Overview of used datasets in the literature}
\end{table*}

\subsection{\textbf{Research Questions}}\label{Research Questions}
This subsection discusses the studies analysis to answer the research questions identified in section~\ref{methodology}.

\textbf{RQ1.} How is AML applied in the NIDS domain? 
\\Based on the surveyed studies, AML has been applied in four areas; generation, evaluation, defense, and synthesis as illustrated in Figure~\ref{distribution of studies per category}. (44\%) of the studies proposed new AEs generation techniques or customized the existing generic AEs generation approaches for crafting adversarial network traffic. (28\%) of the studies utilized the generic AEs generation approaches to assess the resilience of different ML/DL algorithms towards AEs. (26\%) of the studies employed mitigation solutions for AEs such as adversarial training, feature reduction, and ensemble learning. Lastly, AML approaches were utilized by (2\%) of the studies to solve the problem of malicious traffic scarcity and class imbalance in network traffic. 

\textbf{RQ2.} What are the proposed countermeasures to mitigate the threat of AEs against ML-based NIDS?
\\There are no tailored solutions designed for protecting ML-based NIDS against AML attacks. Most of the studies in the category of defending NIDS models against AML attacks employ mitigation countermeasures from other domains such as image recognition. These defensive mechanisms include adversarial training, feature reduction, and ensemble learning. These solutions have their own limitations and do not tackle all attack scenarios. With adversarial training, models' robustness can only be increased to white-box attacks, but they remain vulnerable to black-box attacks, i.e., the AEs generated by other approaches. Feature reduction decreases the detection performance of the models in non-adversarial settings. Ensemble learning increases accuracy on unperturbed examples. This is not the case for other mechanisms as they degrade overall detection performance. The presence of a perturbed input in most applications is an exception; therefore, maintaining high performance on unperturbed inputs while making models more robust to AEs is a critical demand. Moreover, ensemble learning can be easily incorporated with other defensive approaches to enhance the resilience of detection models to AEs. However, ensembles cost more to build, train, and deploy. Their benefits come at the cost of substantial computational expenses and memory requirements.    

\textbf{RQ3.} What are the criteria to assess the ability of adversarial attack generation techniques in producing valid, realistic, and consistent adversarial
flow?
\\Adversarial attacks mainly aim to introduce perturbed examples that force ML-based NIDS models to misclassify decisions and evade detection. However, for realistic attacks, these crafted examples must retain the attack functions and conform to the original characteristics of network traffic. Nevertheless, there are no formal criteria that define network traffic constraints to validate generated adversarial flow. However, any formal description of valid adversarial examples must include inexhaustibly the following~\cite{merzouk2020deeper}: 
\begin{itemize}
    \item limit perturbations to changeable features only. 
    \item introduce values of zero or one for the binary features.
    \item introduce values within the allowed range for the decimal features.
    \item activate only one category at a time for the categorical features.
    \item maintain interdependencies and semantic links between the features.
    \item preserve the attack function of the original flow.
    \item retain the fundamental information of network flow; thus, the direction of perturbation introduced to the original traffic must be strictly confined. 
\end{itemize}

\textbf{RQ4.} What is the feasibility of launching real-world adversarial attacks against ML-based NIDS?
\\Most of the literature assumes threat models improper for real-world scenarios. (60\%) of the studies assume that the adversary has complete knowledge about the targeted detector. In such a threat model, the adversary needs to know the internal configuration of the detector and its decision boundaries. Obtaining such knowledge is unfeasible as the detection system is practically deployed on a highly secure machine accessible to a few devices unlikely to be compromised by the adversary~\cite{apruzzese2021modeling}. On the other hand, some of the surveyed studies assume a black-box setup in which the adversary can reverse engineer the target model to implement a substitute one by sending some crafted inputs and observing the corresponding outputs. In this situation, the adversary must limit the number of queries to avoid triggering the detection mechanism and operate in a low-and-slow approach which extends the attack campaign up to days or weeks~\cite{apruzzese2021modeling}. Moreover, the adversary cannot directly observe the feedback outside of improbable circumstances, such as the NIDS employing a reactive mechanism, gaining access to the NIDS's log or console, or acquiring the same commercial NIDS and interacting with it freely in a controlled environment.   

\subsection{\textbf{Findings Analysis}}
\begin{figure}[htbp]
 \centering
 \includegraphics[scale=0.35]{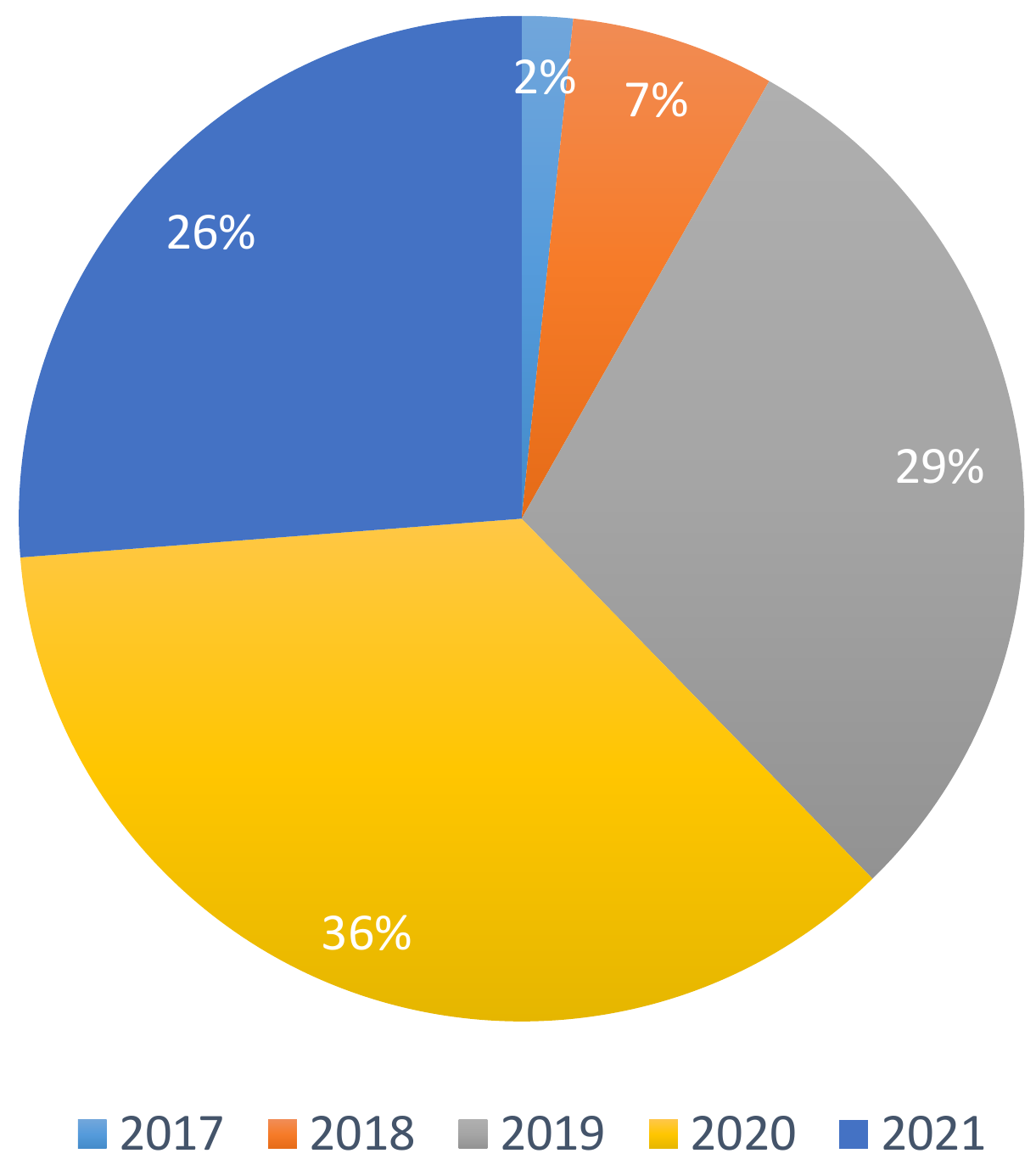}
 \caption{distribution of studies per year}
\label{distribution of studies per year}
\end{figure}

Growing interest in the applications of adversarial machine learning applications into the network intrusion detection domain is demonstrated by the gradual yearly increase of publications, shown in figure~\ref{distribution of studies per year}. 

\begin{figure}[htbp]
 \centering
 \includegraphics[scale=0.35]{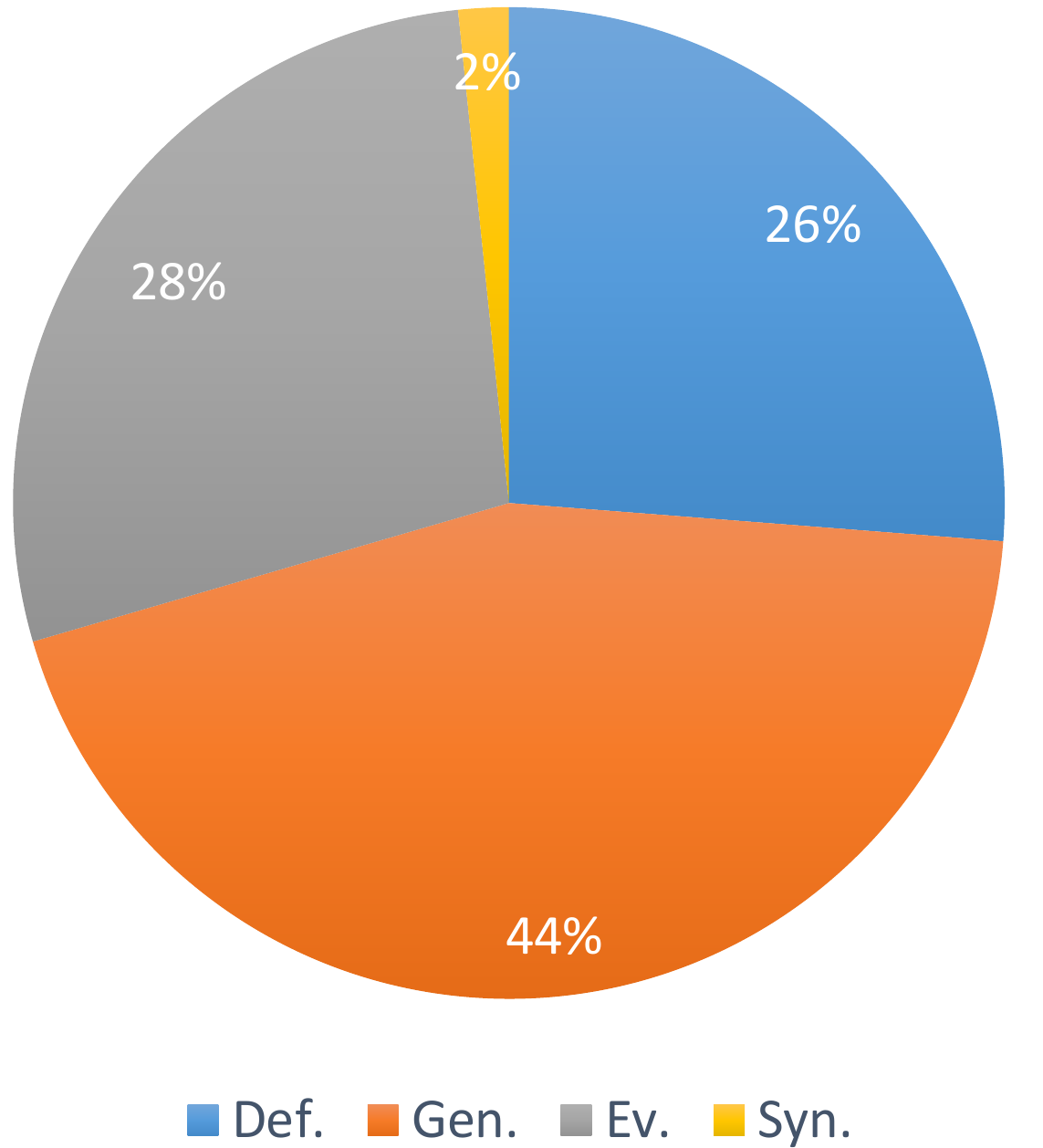}
 \caption{distribution of studies per category}
\label{distribution of studies per category}
\end{figure}

As can be seen in figure~\ref{distribution of studies per category}, the majority of the studies (44\%) fall in the category of proposing new techniques for generating adversarial attacks against NIDs, followed by evaluating the robustness of ML-based NIDS towards adversarial examples generated by generic approaches (28\%). These generation approaches showed significant degradation of NIDS's performance, which indicates the highly risky implications of adversarial examples on network security. Notably, fewer studies (26\%) proposed new defensive solutions or evaluated the suitability of some existing defense mechanisms for NIDS. It is noteworthy that these proposed solutions are mainly limited to adversarial training, ensemble learning, and feature reduction. Each of these techniques suffers considerable shortcomings makes it ineffective for protecting NIDSs from adversarial attacks. 

\begin{figure}[htbp]
 \centering
 \includegraphics[scale=0.35]{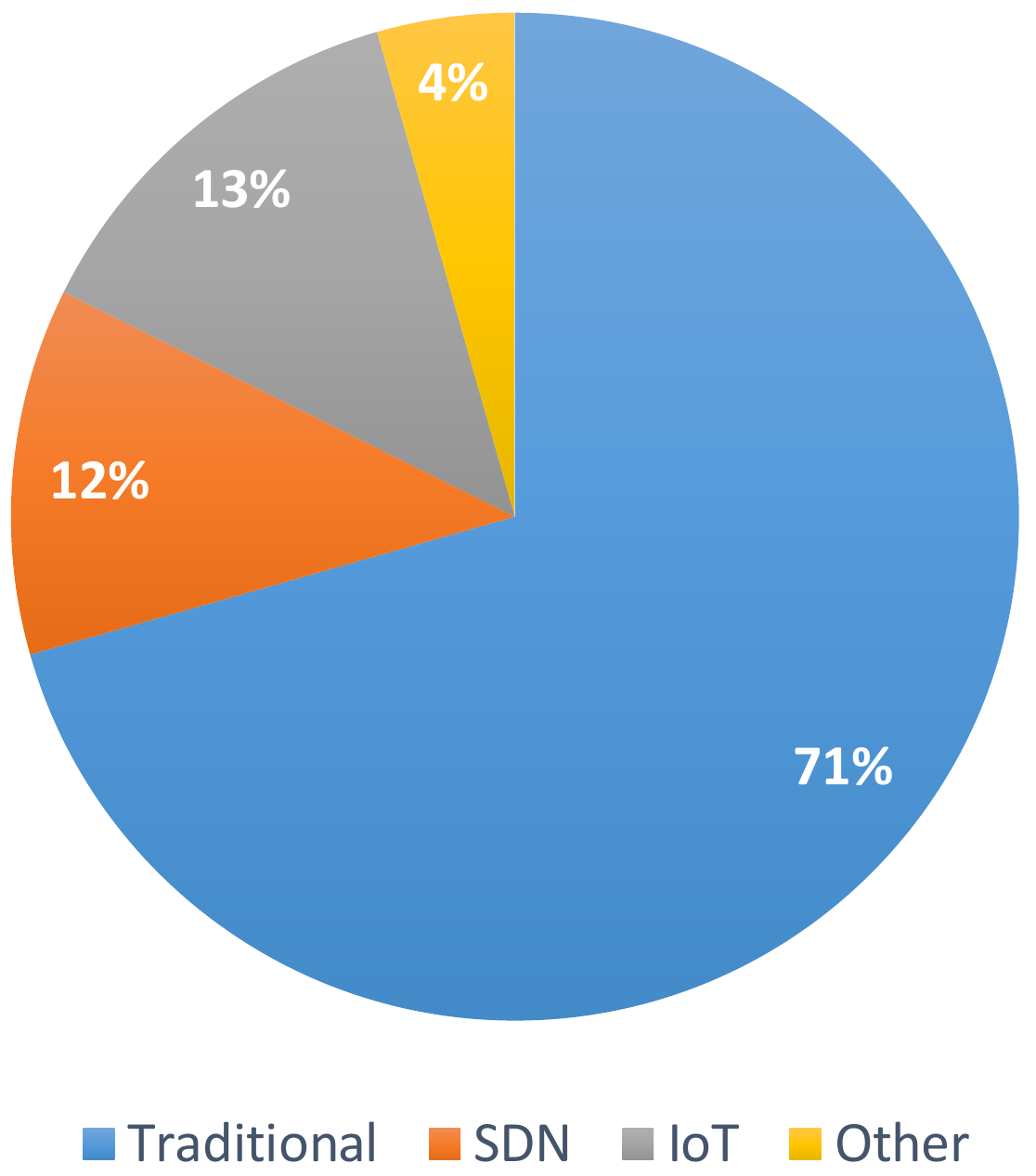}
 \caption{distribution of studies per environment}
\label{distribution of studies per enviroment}
\end{figure}

Figure~\ref{distribution of studies per enviroment} shows the distribution of papers addressing different types of networks. It can be seen that most of the studies focused on traditional networks (71\%), while one study dealt with wireless networks (2\%). Furthermore, fewer studies investigated adversarial attacks in IoT networks (13\%). Such networks emerge in various contexts (e.g., smart homes, smart buildings, etc.) and demand protection from adversarial attacks. 

\begin{figure}[htbp]
 \centering
 \includegraphics[scale=0.35]{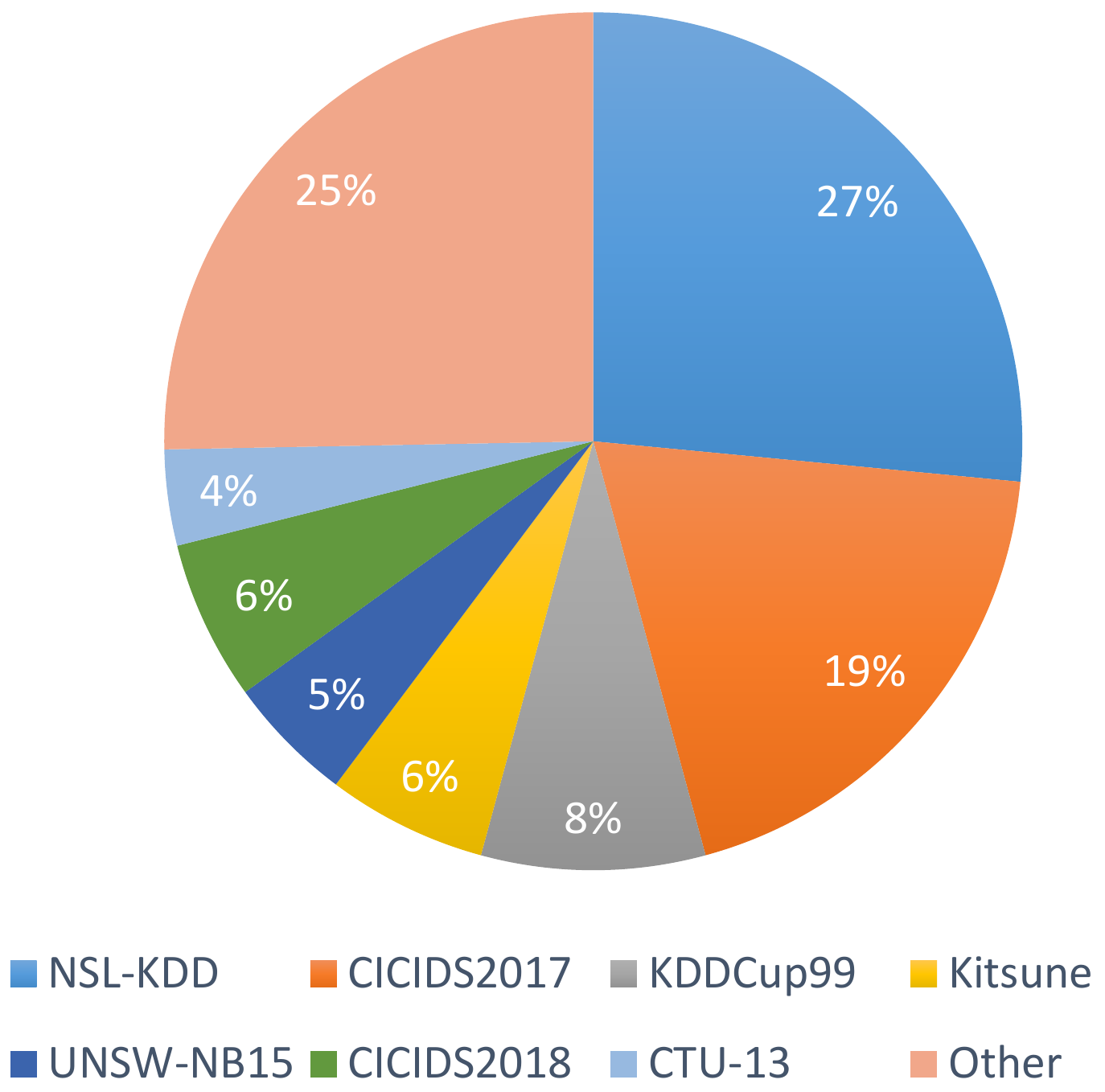}
 \caption{distribution of studies per dataset}
\label{distribution of studies per dataset}
\end{figure}

As shown in figure~\ref{distribution of studies per dataset}, the NSL-KDD, an enhanced version of KDDCup99, is the most used dataset by (27\%) of the surveyed studies. This dataset is ideal, outdated, and is not representative of today's networks. Similarly, the KDDCup99 dataset, used by (8\%) of the studies, suffers the same shortcomings along with containing a large portion of redundant instances that affect models' learning process by inducing the ML algorithms to be biased towards the frequently duplicated records~\cite{tavallaee2009detailed}. Recent datasets that reflect the complexity of real-worlds networks need to be adapted for evaluating the performance of NIDSs under adversarial attacks. Moreover, there is a need to utilize commonly standard datasets for benchmarking purposes as (25\%) of the studies either used uncommon datasets or emulated their ones.  

\begin{figure}[htbp]
 \centering
 \includegraphics[scale=0.35]{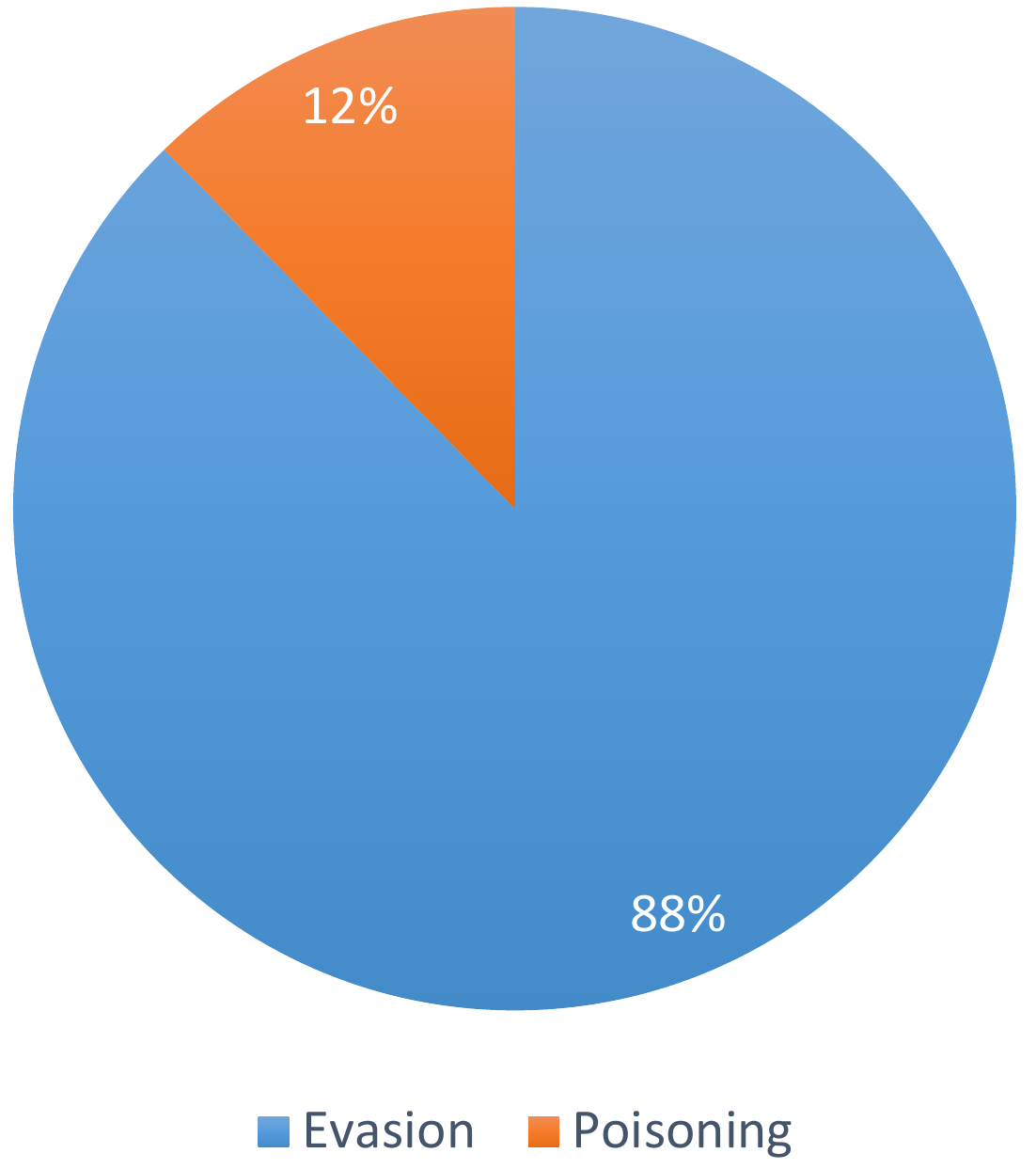}
 \caption{distribution of studies per strategy}
\label{distribution of studies per startegy}
\end{figure}

Figure~\ref{distribution of studies per startegy} shows a lack of studies (12\%) addressing adversarial poisoning attacks. These attacks are typically untargeted and aim to degrade the model's overall detection accuracy on the clean examples. They are also often called availability attacks as they yield to decrease the model's accuracy, akin to denial of service attacks. The adversary injects malicious inputs during the training phase, influencing the models' accuracy after the deployment. These inputs have similar features but with wrong labels, which disturbs the training data distribution. Poisoning attacks aim to shift the model's decision boundaries and cause concept drift (i.e., changing statistical properties of predicted variables in unforeseen means overtime). These attacks are achieved by infusing noise to the training inputs before feature extraction, altering inputs labels, or crafting synthetic inputs to induce slow concept drift. Although the training data's integrity and confidentiality are more likely to be protected in practice, the adversary can exploit this vulnerability during retraining the existing models. The NIDS models require periodic updating to adapt to changing contexts. Such scenarios demand online training to update the learned model with new incoming training data, and poisoning attacks emerge as the major threat to these systems during retraining procedures. For instance, the attacker can compromise a few devices in the network (e.g., sensors) to submit malicious traffic, which is then used to retrain the detection model and cause subverting the learning process.  

Furthermore, these attacks present critical threats to Federated Learning (FL), which one of its important deployment fields is emerging IoT networks. FL enables deploying ML applications on large scales and provides advantages such as preserving users' privacy and efficient service delivery.  Recently, FL has been utilized for IoT network intrusion detection, where a local gateway monitors traffic of IoT devices on its network, trains a local detection model, and then sends this model to a central entity (i.e., aggregator)~\cite{nguyen2020poisoning}. The aggregator uses a federated averaging algorithm to construct a global detection model based on the received models from all gateways and then propagates this model to the local gateways. This distributed training mechanism of FL protects users' privacy as no need for sharing the local data with other parties. Additionally, aggregating local models enables building an accurate detection model quickly, especially for the devices that generate little data. However, FL can be targeted by a backdoor attack, a type of poisoning attack in which the adversary gradually injects malicious inputs using compromised IoT devices to corrupt the resulting aggregated model. The current techniques are ineffective in defending the DL-based models from poisoning attacks. The statistical-based approaches can enhance models' resilience to noise but perform poorly on adversarial poisoning examples, and data sanitization techniques operate under restrictive assumptions.  

\begin{figure}[htbp]
 \centering
 \includegraphics[scale=0.35]{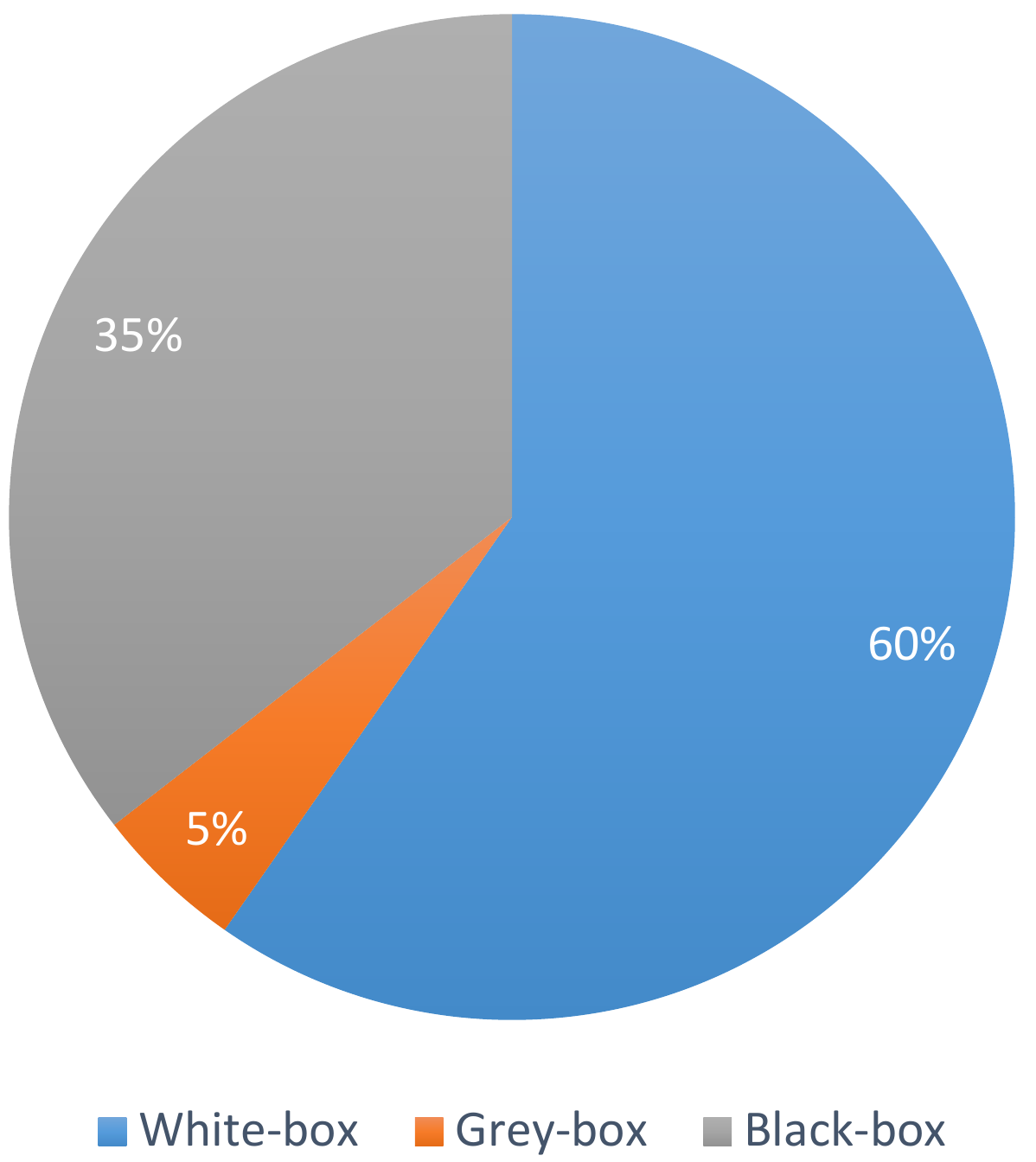}
 \caption{distribution of studies per setting}
\label{distribution of studies per setting}
\end{figure}

Figure~\ref{distribution of studies per setting} illustrates that a considerable proportion of the studies addressed white-box attacks (60\%). Although such attacks are unlikely practical in real scenarios, they help the designers to assess models' resilience to adversarial examples. The experimental results of black-box studies showed remarkable evasion rates, which indicates high potentials for adversaries to compromise NIDS without any prior knowledge.
 \begin{figure}[htbp]
 \centering
 \includegraphics[scale=0.35]{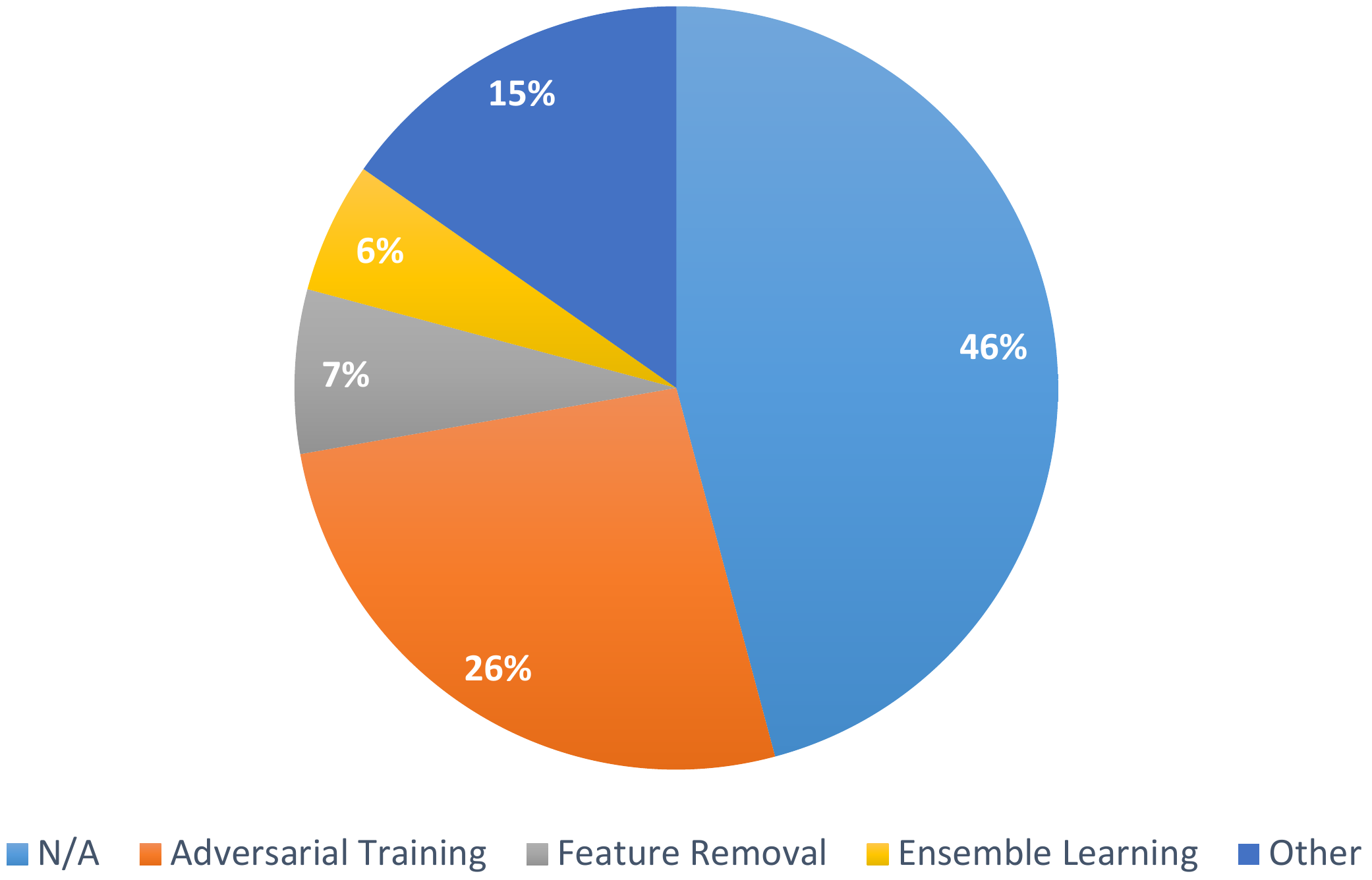}
 \caption{distribution of studies per defense mechanism}
\label{distribution of studies per defense}
\end{figure}

Figure~\ref{distribution of studies per defense} demonstrates a drastic lack of mitigation solutions for adversarial attacks as the majority of the studies (46\%) did not address that. The major solutions that have been investigated in the literature are limited to adversarial training (26\%), feature removal (7\%), and ensemble learning (6\%). Adversarial training is inefficient against attacks that differ from the ones in the training dataset. Feature removal reduces the detection model's performance in the non-adversarial setting, and ensemble learning suffers from a lack of interpretability and demands expensive computations.

\begin{figure}[htbp]
 \centering
\includegraphics[scale=0.35]{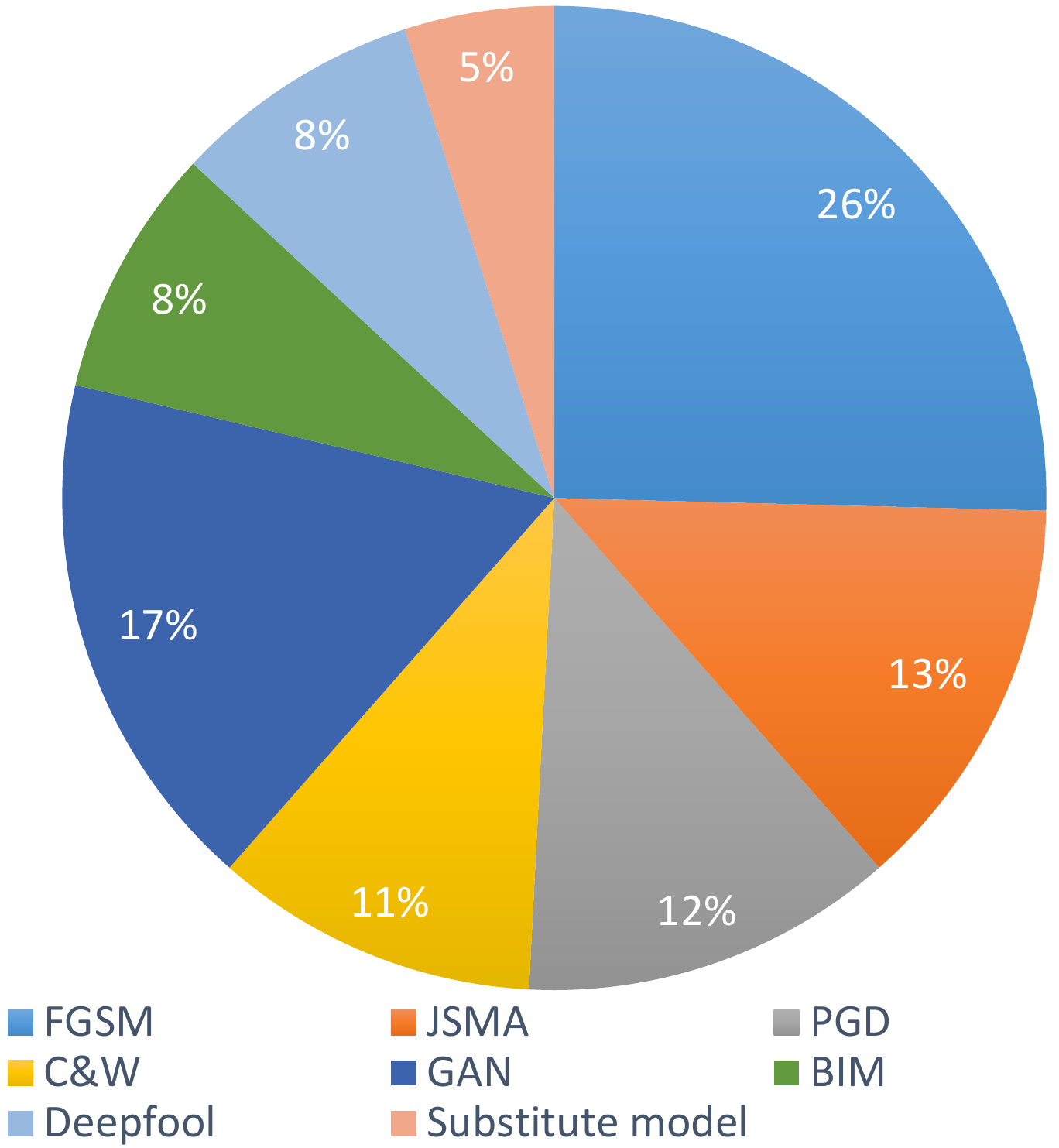}
 \caption{Most utilized generic adversarial attacks}
\label{Most utilized generic adversarial attacks}
\end{figure}

Figure~\ref{Most utilized generic adversarial attacks} shows the most utilized generic approaches for crafting adversarial examples to evade NIDSs. The utilization frequency of these approaches is as follows: FGSM (17\%), JSMA (13\%), C\&W (11\%), GAN (17\%), PGD (12\%), DeepFool (8\%), BIM (8\%). FGSM is the most utilized approach as it is one of the easiest and yet effective methods. However, it is practically not a suitable method for the NIDS domain as it changes each feature to produce adversarial examples. It is unlikely for the adversary to have complete control to change all features in such a fine-grained manner. Moreover, network traffic features are correlated, highly interdependent, and domain-constrained. Additionally, most of these utilized attacks are white-box that work under the threat model of a strong adversary with some knowledge about the targeted system. Although such a setting is not practically common, a weaker adversary with only access to the targeted system's outputs and some knowledge about the expected inputs can utilize the model as an oracle. The adversary can create a substitute model using a limited number of synthetic inputs and their corresponding outputs (i.e., confidence prediction scores returned by the victim model). Due to the transferability of AEs, the adversary crafts examples to evade the substitute model, and these examples are then used to bypass the targeted system.

Although many studies have examined the generic adversarial attacks for ML-based NIDS, the consistency and validity of the generated AEs by these approaches were not addressed. These AEs can fool the detection models; however, they cannot be practically implemented for end-to-end attacks. 

Network traffic features differ from other domains such as image recognition, where the features are unstructured and unconstrained. Network features are constrained by specific data types and ranges of values.  Each feature can have a binary, categorical, or continuous value, which is limited to a particular range. The generic approaches do not maintain these constraints and result in invalid and inconsistent adversarial traffic. For instance, FGSM introduces non-binary values to the binary features in network traffic. Instead of changing binary features from one state to another, FGSM adds 0.1 as an applied perturbation~\cite{10.1007/978-3-030-68887-5_4}. Some network traffic features contain label values rather than numeric values (i.e., categorical or nominal features). Many ML algorithms cannot work with label values directly and require all the values to be numeric. Therefore, categorical features are converted to multiple binary features using one-hot encoding. Only one binary feature can take the value 1, while the rest must take the value 0.  However, because FGSM propagates perturbations over all the features, that results in activating multiple categories at the same time~\cite{10.1007/978-3-030-68887-5_4}. Similarly, the other approaches (i.e., BIM, DeepFool, C\&W, and JSMA) inherit the same flaws.

\begin{figure}[htbp]
\centering
\includegraphics[scale=0.40]{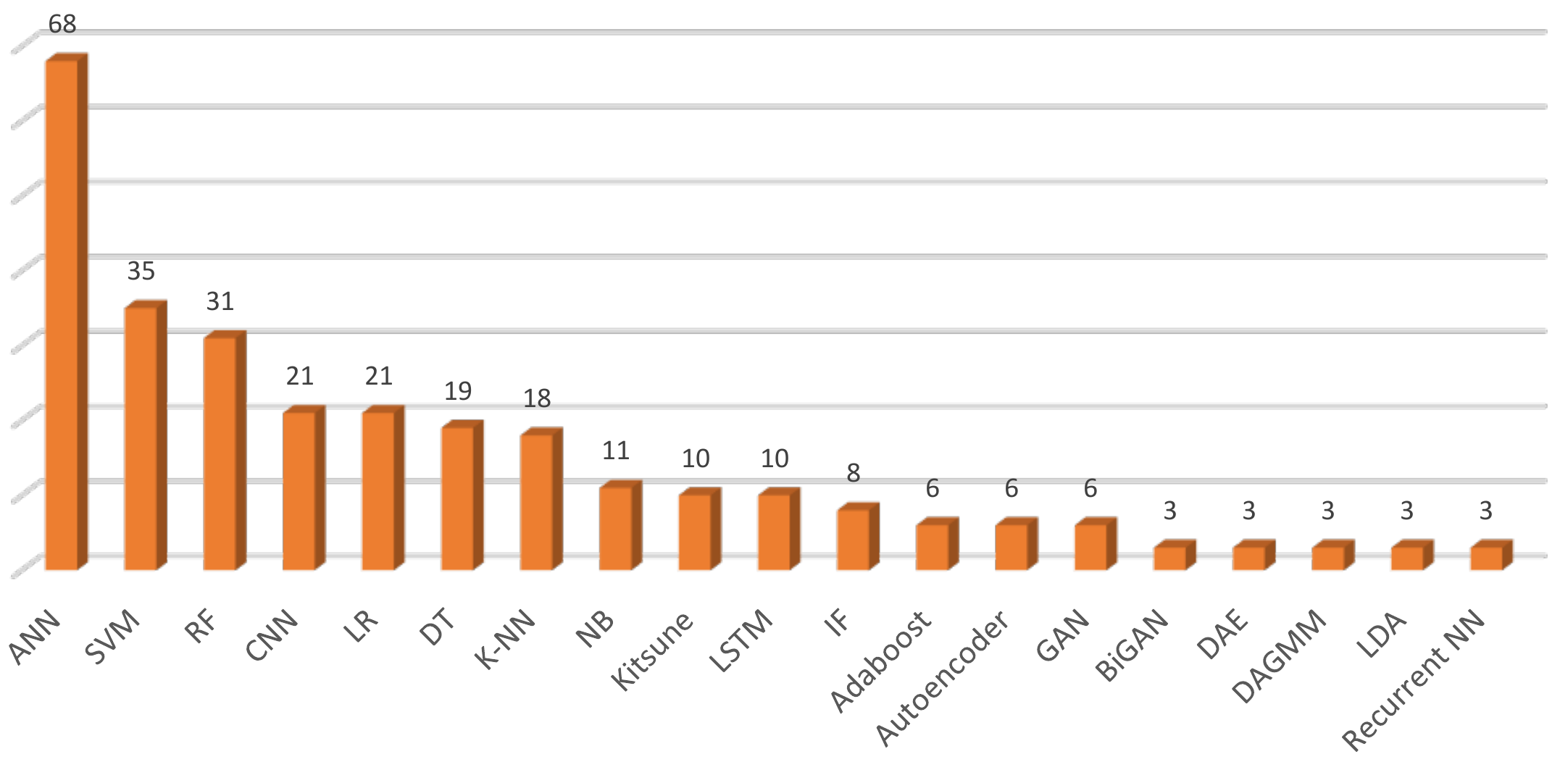}
 \caption{distribution of attacked models}
\label{Distribution of attacked models}
\end{figure}

Figure~\ref {Distribution of attacked models} presents the distribution of attacked models among the surveyed studies. A wide variety of conventional-based and deep-based NIDS models were utilized to assess their resilience against AEs generated using different generic and customed approaches. Overall, these well-known NIDS models showed similar performance degradation under adversarial attacks.

 The remarkable notes concluded from the investigated studies as follows: 
 
 \begin{itemize}
    \item NIDSs operate in adversarial environments; hence the threat of adversarial examples towards such systems is more severe. Thus, there is an immediate demand to design robust NIDSs that are resilient to adversarial attacks. 
    \item Adversarial attacks can significantly degrade the performance of NIDS classifiers. Due to the transferability of AEs across different architectures of ML-based NIDS, the surface attack of such systems is increasingly expanding. 
    \item A wide variety of adversarial attack generation techniques have been utilized. However, it is noticeable that most of these approaches are white-box based (60\%). Such a threat model is not commonly feasible for adversaries. Black-box attacks are more practical in real-world scenarios, and their threat and defense against them demand to be addressed.    
    \item The majority of studies did not take into consideration the domain constraints on NIDS features. Unlike unconstrained domains (e.g., image recognition and natural language processing), in the constrained ones (e.g., NIDS), the dataset features have the following characteristics;  each feature can have a binary, categorical, or continuous value, the values of different features are correlated and highly-interdependent, some features are fixed and unmodifiable. The adversarial generation techniques need to maintain these conditions and craft a realistic traffic flow that can evade the detection system. 
    \item The feasibility of launching these attacks in real-world scenarios has not been addressed in all studies. Many factors constrain selecting the appropriate adversarial generation technique, such as needed computation power, number of modifiable features, and magnitude of introduced perturbations.   
    \item There is a noticeable lack of research studies (46\%) that either introduce new or investigate the effectiveness of existent adversarial defensive mechanisms for the NIDS domain. The proposed solutions include adversarial training, feature removal, and ensemble learning. However, other defensive mechanisms have been proposed in the image recognition domain, and their applicability and effectiveness in the area of NIDS have not been explored with the need to consider the unique challenges and differences of the NIDS domain. For instance, the overhead of the defensive mechanism is critical as NIDS classifiers usually operate in real-time, demanding a low overhead method, unlike in the image classification domain. Notably, although such generic countermeasures exist, none of them can address all challenges. The attention should be paid to introducing attack-agnostic defense approaches that tackle the increasing variety of adversarial attacks, not to the attack-specific that are designed only for a narrow range of attacks such as adversarial training. Therefore, devising a highly resilient design for machine learning applications against adversarial attacks remains an open research problem.  
    \item The majority of the studies have utilized exclusively the greatly outdated NSL-KDD dataset (35\%). This dataset suffers significant shortcomings and is not an ideal representative of existing real-world networks. Therefore, there is a need to adapt the use of more recent and standardized NIDS datasets to address emerging threats by modern technologies such as IoT networks.
    \item Most of the utilized defensive techniques, their effectiveness was tested on white-box attacks. To address the emerging challenges by the black-box adversaries, these defensive methods must be assessed comprehensively in both settings to increase their applicability and robustness for a wide variety of scenarios.    
\end{itemize}
  \label{discussion}
\section{Conclusion} Despite their exalted performance and efficiency, DL-based NIDSs are vulnerable to crafted tiny perturbations that are added to the legitimate traffic in order to deceive the detection systems, which causes catastrophic consequences to the network security. Such an aspect of DL-NIDS demands immediate attention as it has not been relatively investigated. This study presents a comprehensive view of adversarial machine learning attacks applied to the network anomaly detection domain. It categorizes the surveyed studies based on their contribution mainly into three groups: generating AEs to evade ML-based NIDSs, evaluating their robustness, and defending them against adversarial attacks. Furthermore, it discusses the shortcomings of the existing defensive mechanisms, the applicability of the existing adversarial approaches for implementing successful attacks, and the feasibility of launching these attacks in real-world scenarios. Overall, this study extensively identifies, evaluates, and synthesizes the existing literature to provide a clear view, identify research gaps, and conclude remarkable notes for future work.

 \label{conclusion}
\onecolumn
\footnotesize
\begin{longtable}{|l|l|l|P{1.5cm}|P{1.5cm}|P{1.5cm}|P{1.5cm}|P{2cm}|P{2cm}|P{2cm}|}
\endfoot
\hline
Category	&	Ref.	 & 	Year	 & 	Environment	 & 	Dataset	 &	Setting	 & 	Strategy	 & 	Attack	 & 	Models	 & 	Defense	\\ \hline
\endhead
	&	~\cite{usama2019generative}	 & 	2019	 & 	Traditional 	 & 	KDDCUP99	 &	Black-box	 & 	Evasion	 & 	GAN  	 & 	DNN, GAN, K-NN, RF, SVM	 & 	GAN-based Adversarial Training	\\ \cline{2-10}
	&	~\cite{zhang2020tiki}	 & 	2020	 & 	Traditional 	 & 	CICIDS2018	 &	Black-box	 & 	Evasion	 & 	Boundary, HSJA, NES, Opt, Pointwise	 & 	CNN, LSTM, MLP	 & 	Adversarial Query Detection, Adversarial Training, Ensemble Learning	\\ \cline{2-10}
	&	~\cite{abou2020evaluation}	 & 	2020	 & 	Traditional 	 & 	NSL-KDD, UNSW-NB15	 &	White-box	 & 	Evasion	 & 	BIM, C\&W, Deepfool, FGSM, PGD	 & 	DNN, CNN, Recurrent NN	 & 	Min-Max Formulation	\\ \cline{2-10}
	&	~\cite{AbouKhamis2019}	 & 	2019	 & 	Traditional 	 & 	UNSW-NB15	 &	White-box	 & 	Evasion	 & 	BCAS, BGAS, dFGSMS, rFGSMS	 & 	DNN	 & 	Min-Max Formulation, Adversarial Training, Feature Reduction (i.e., PCA)	\\ \cline{2-10}
Def.	&	~\cite{Qureshi2020}	 & 	2020	 & 	Traditional 	 & 	NSL-KDD	 &	White-box	 & 	Evasion, Poisoning	 & 	 JSMA	 & 	RNN-ADV trained with ABC, DNN	 & 	RNN-ADV trained with ABC	\\ \cline{2-10}
	&	~\cite{Apruzzese2020}	 & 	2020	 & 	Traditional 	 & 	CTU-13	 &	Grey-box	 & 	Evasion	 & 	~\cite{apruzzese2018evading}	 & 	Adaboost, DT, MLP, RF, WnD	 & 	Queries attempts restriction, Ensemble Learning	\\ \cline{2-10}
	&	~\cite{hashemi2020enhancing}	 & 	2020	 & 	Traditional 	 & 	CICIDS2017	 &	White-box	 & 	Evasion	 & 	~\cite{hashemi2019towards}	 & 	DAE, Kitsune, BiGAN, DAGMM	 & 	RePO	\\ \cline{2-10}
	&	~\cite{benzaid2020robust}	 & 	2020	 & 	SDN 	 & 	CICIDS2017	 &	White-box	 & 	Poisoning	 & 	FGSM 	 & 	MLP	 & 	Adversarial Training	\\ \cline{2-10}
	&	~\cite{pawlicki2020defending}	 & 	2020	 & 	Traditional 	 & 	CICIDS2017	 &	White-box	 & 	Evasion	 & 	BIM, C\&W, FGSM, PGD	 & 	Adaboost, DNN, RF, SVM, K-NN	 & 	Neurons Activation	\\ \cline{2-10}
	&	~\cite{debicha12detect}	&	2021	&	Traditional	&	NSL-KDD	&	White-box	&	Evasion	&	FGSM, PGD	&	DNN, DT, LDA, LR, RF, SVM	&	Ensemble Learning, Detect \& Reject	\\ \cline{2-10}	
    &	~\cite{ganesan2021mitigating}	&	2021	&	SDN	&	CICIDS2017, DARPA, KDDCup99	&	White-box	&	Evasion	&	Hydra	&	DNN, LR, RF, SVM	&	Feature Removal, Ensemble Learning	\\ \cline{2-10}	
    &	~\cite{mccarthy2021feature}	&	2021	&	Traditional	&	CICIDS2017	&	White-box	&	Evasion	&	FGSM	&	DNN	&	Feature Removal	\\ \cline{2-10}	
    &	~\cite{novaes2021adversarial}	&	2021	&	Traditional, SDN	&	CICIDS2019, Emulated Real SDN environment	&	White-box	&	Evasion	&	GAN	&	CNN, GAN, LSTM, MLP	&	GAN-based Adversarial Training	\\ \cline{2-10}	
    &	~\cite{nugraha2021detecting}	&	2021	&	SDN	&	Theirs, Emulated Real SDN environment	&	White-box	&	Evasion	&	Tabular GAN, Emulated Adversary SDN dataset	&	MLP, CNN-LSTM	&	Adversarial Training	\\ \cline{2-10}	
    &	~\cite{wang2021def}	&	2021	&	Traditional	&	CICIDS2018	&	White-box	&	Evasion	&	BIM, Deepfool, FGSM, JSMA	&	DNN	&	Ensemble Retraining	\\ \cline{2-10}	

	&	~\cite{peng2020detecting}	 & 	2020	 & 	Traditional 	 & 	NSL-KDD	 &	White-box	 & 	Evasion	 & 	FGSM, MIFGSM, PGD	 & 	DNN	 & 	BiGAN-based Adversarial Training	\\ \hline
	&	~\cite{lin2018idsgan}	 & 	2018	 & 	Traditional 	 & 	NSL-KDD	 &	Black-box	 & 	Evasion	 & 	WGAN 	 & 	DT, K-NN, LR, MLP, NB, RF, SVM	 & 	N/A	\\ \cline{2-10}
	&	~\cite{Li2018}	 & 	2018	 & 	Traditional, Wireless 	 & 	NSL-KDD, Kyoto2006+, WSN-DS	 &	Black-box	 & 	Poisoning, Stealing Model	 & 	A-SMOTE, CDBP	 & 	DNN	 & 	N/A	\\ \cline{2-10}
	&	~\cite{kuppa2019black}	 & 	2019	 & 	Traditional 	 & 	CICIDS2018	 &	Black-box	 & 	Evasion	 & 	Manifold Approximation Algorithm, spherial local subspaces	 & 	AE, BiGAN, DAGMM, GAN, IF, One-Class SVM	 & 	N/A	\\ \cline{2-10}
	&	~\cite{Yan2019}	 & 	2019	 & 	Traditional 	 & 	KDDCup99	 &	Black-box	 & 	Evasion	 & 	WGAN	 & 	CNN	 & 	N/A	\\ \cline{2-10}
	&	~\cite{Abusnaina2019a}	 & 	2019	 & 	SDN 	 & 	~\cite{niyaz2016deep}	 &	White-box	 & 	Evasion	 & 	C\&W, Deepfool, ENM, MIM, PGD, FlowMerge	 & 	CNN	 & 	Adversarial Training	\\ \cline{2-10}
Gen.	&	~\cite{peng2019adversarial}	 & 	2019	 & 	Traditional 	 & 	CICIDS2017, KDDCUP99	 &	Black-box	 & 	Evasion, Poisoning	 & 	Improved Boundary-Based method	 & 	DNN	 & 	N/A	\\ \cline{2-10}
	&	~\cite{usama2019black}	 & 	2019	 & 	Tor 	 & 	UNB-CIC Tor	 &	Black-box	 & 	Evasion	 & 	MI, Substitute Model	 & 	DNN, SVM	 & 	N/A	\\ \cline{2-10}
	&	~\cite{aiken2019investigating} 	 & 	2019	 & 	SDN 	 & 	CICIDS2017, DARPA	 &	Black-box	 & 	Evasion	 & 	Manipulating three traffic features	 & 	RF, SVM, LR, K-NN	 & 	N/A	\\ \cline{2-10}
	&	~\cite{usama2019adversarial}	 & 	2019	 & 	Traditional 	 & 	NSL-KDD	 &	White-box	 & 	Evasion	 & 	MI, minimizing L1 between discriminating features of normal \& DoS	 & 	DNN, SVM	 & 	N/A	\\ \hline
	&	~\cite{Zhang2020}	 & 	2020	 & 	Linux HIDS, Anroid Malware, Traditional 	 & 	ADFA-LD, DREBIN, NSL-KDD	 &	Black-box	 & 	Evasion	 & 	BFAM	 & 	LR, MLP, NB, RF	 & 	N/A	\\ \cline{2-10}
	&	~\cite{han2020practical}	 & 	2020	 & 	IoT, Traditional 	 & 	Kitsune, CICIDS2017	 &	Black-box \& Grey-box	 & 	Evasion	 & 	GAN, PSO	 & 	Kitsune, MLP, LR, DT, SVM, IF	 & 	Feature Removal	\\ \cline{2-10}
	&	~\cite{Yang2020}	 & 	2020	 & 	Traditional 	 & 	KDDCUP99	 &	White-box	 & 	Evasion	 & 	U-ASG	 & 	Deterministic AE, VAE	 & 	N/A	\\ \cline{2-10}
	&	~\cite{wang2020c}	 & 	2020	 & 	Traditional  	 & 	NSL-KDD	 &	White-box	 & 	Evasion	 & 	CIFGSM, IFGSM	 & 	DT, CNN, MLP	 & 	N/A	\\ \cline{2-10}
	&	~\cite{qiu2020adversarial}	 & 	2020	 & 	IoT 	 & 	Kitsune	 &	Black-box	 & 	Evasion	 & 	IFGSM, Substitute Model, Saliency Maps	 & 	Kitsune	 & 	N/A	\\ \cline{2-10}
	&	~\cite{khamaiseh2020deceiving}	 & 	2020	 & 	SDN 	 & 	simualted	 &	Black-box	 & 	Evasion	 & 	Customed purebations	 & 	DNN, IF, K-NN, NB, SVM	 & 	N/A\\	\cline{2-10}
	&	~\cite{alhajjar2020adversarial}	 & 	2020	 & 	Traditional 	 & 	NSL-KDD, CICIDS2017	 &	White-box	 & 	Evasion	 & 	PSO, GA, GAN	 & 	SVM, DT, NB, K-NN, RF, MLP, GB, LR, LDA, QDA, BAG	 & 	N/A	\\ \cline{2-10}
	&	~\cite{shu2020generative}	 & 	2020	 & 	Traditional 	 & 	CICIDS2017	 &	Black-box	 & 	Evasion	 & 	GAN with active learning 	 & 	Gradient Boosted DT	 & 	N/A	\\ \cline{2-10}
Gen.	&	~\cite{chauhan2020polymorphic}	 & 	2020	 & 	Traditional 	 & 	CICIDS2017	 &	Black-box	 & 	Evasion, Poisoning	 & 	GAN, Updating the feature profile	 & 	DT, LR, NB, RF	 & 	Adversarial Training	\\ \cline{2-10}
	&	~\cite{Teuffenbach2020}	 & 	2020	 & 	Traditional 	 & 	CICIDS2017, NSL-KDD	 &	White-box	 & 	Evasion	 & 	C\&W optimizes based on perturbation constraints \& feature weights	 & 	AE, DBN, DNN	 & 	N/A	\\ \cline{2-10}
	&	~\cite{chen2020fooling}	 & 	2020	 & 	Traditional 	 & 	CICIDS2017, NSL-KDD, UNSW-NB15	 &	White-box	 & 	Evasion	 & 	AE, GAN  	 & 	Adaboost, CNN, DT, K-NN, LR, LSTM, RF	 & 	N/A	\\ \cline{2-10}
	&	~\cite{shiehdetection}	&	2021	&	Traditional	&	NSL-KDD	&	White-box	&	Evasion	&	WGAN with gradient penalty	&	K-NN, MLP, RF	&	GAN-based Adversarial Training	\\ \cline{2-10}		
    &	~\cite{cheng2021packet}	&	2021	&	Traditional	&	CTU-13	&	Black-box	&	Evasion	&	Seq-GAN	&	DT, LR, MLP, SVM	&	N/A	\\ \cline{2-10}		
    &	~\cite{gomez2021crafting}	&	2021	&	ICS	&	Electra	&	White-box	&	Evasion	&	Selective and Iterative Gradient Sign Method	&	Dense Neural Network	&	N/A	\\ \cline{2-10}		&	~\cite{guo2021black}	&	2021	&	Traditional	&	CICIDS2018, KDDCup99	&	Black-box	&	Evasion	&	BIM, Substitute model	&	CNN, K-NN, MLP, SVM, Residual Network	&	N/A	\\ \cline{2-10}		
    &	~\cite{anthi2021hardening}	&	2021	&	IoT	&	Smart Home IoT testbed	&	White-box	&	Evasion	&	InfoGain Ratio, Feature Importance	&	RF, SVM, J48 Decision Tree, Bayesian Network	&	Adversarial Training	\\ \cline{2-10}		
    &	~\cite{sharon2021tantra}	&	2021	&	Traditional, IoT	&	CICIDS2017, Kitsune	&	Black-box	&	Evasion	&	Timing-Based Adversarial Network Traffic Reshaping Attack, LSTM	&	Autoencoder, IF, Kitsune	&	Adversarial Training	\\ \cline{2-10}		
	&	~\cite{apruzzese2019addressing}	 & 	2019	 & 	Traditional	 & 	CTU-13	 &	Black-box	 & 	Evasion, Poisoning	 & 	Adding random values within feature's interval	 & 	K-NN, MLP, RF	 & 	Adversarial Training, Feature Removal	\\ \hline
	&	~\cite{rigaki2017adversarial}	 & 	2017	 & 	Traditional 	 & 	NSL-KDD	 &	Grey-box	 & 	Poisoning	 & 	JSMA	 & 	DT, MLP, RF, SVM, Majority Voting Ensemble	 & 	N/A	\\ \cline{2-10}
	&	~\cite{warzynski2018intrusion}	 & 	2018	 & 	Traditional 	 & 	NSL-KDD	 &	White-box	 & 	Evasion	 & 	FGSM 	 & 	DNN	 & 	N/A	\\ \cline{2-10}
	&	~\cite{Wang2018}	 & 	2018	 & 	Traditional 	 & 	NSL-KDD	 &	White-box	 & 	Evasion	 & 	C\&W, Deepfool, FGSM, JSMA	 & 	MLP	 & 	N/A	\\ \cline{2-10}
Eva.	&	~\cite{Yang2019}	 & 	2019	 & 	Traditional 	 & 	NSL-KDD	 &	Black-box	 & 	Evasion	 & 	WGAN, Substitute Model, Zoo	 & 	DNN	 & 	N/A	\\ \cline{2-10}
	&	~\cite{clements2019rallying} 	 & 	2019	 & 	IoT 	 & 	Kitsune	 &	White-box	 & 	Evasion	 & 	FGSM, JSMA, C\&W, ENM	 & 	Kitsune	 & 	N/A	\\ \cline{2-10}
	&	~\cite{Huang2019}	 & 	2019	 & 	SDN 	 & 	Theirs	 &	White-box	 & 	Evasion	 & 	FGSM, JSMA, JSMA-RE	 & 	CNN, LSTM, MLP	 & 	N/A	\\ \cline{2-10}
	&	~\cite{Ibitoye2019}	 & 	2019	 & 	IoT 	 & 	BoT-IoT	 &	White-box	 & 	Evasion	 & 	BIM, FGSM, PGD	 & 	DNN, SNN	 & 	Feature Normalization	\\ \cline{2-10}
	&	~\cite{martins2019analyzing}	 & 	2019	 & 	Traditional 	 & 	CICIDS2017, NSL-KDD	 &	White-box	 & 	Evasion, Poisoning	 & 	C\&W, Deepfool, FGSM, JSMA	 & 	DT, DNN, DAE, NB, RF, SVM	 & 	Adversarial Training	\\ \cline{2-10}
	&	~\cite{Peng2019}	 & 	2019	 & 	Traditional 	 & 	NSL-KDD	 &	White-box	 & 	Evasion	 & 	L-BFGS, MIFGSM, PGD, SPSA	 & 	DNN, LR, RF, SVM	 & 	N/A	\\ \cline{2-10}
	&	~\cite{jeong2019adversarial}	 & 	2019	 & 	Traditional 	 & 	NSL-KDD	 &	White-box	 & 	Evasion	 & 	FGSM, JSMA	 & 	AE, CNN	 & 	N/A	\\ \cline{2-10}
	&	~\cite{Sriram2020}	 & 	2020	 & 	Traditional 	 & 	NSL-KDD	 &	White-box	 & 	Evasion	 & 	FGSM, JSMA	 & 	Adaboost, CNN, DT, DNN, K-Means, LR, LSTM, NB, RF, SVM	 & 	N/A	\\ \cline{2-10}
	&	~\cite{zhong2020adversarial}	 & 	2020	 & 	Traditional, IoT 	 & 	CICIDS2017, Kitsune	 &	Black-box	 & 	Evasion	 & 	WGAN 	 & 	IF, Kitsune, RBM, SAE, SVM	 & 	N/A	\\ \cline{2-10}
	&	~\cite{pacheco2021adversarial}	&	2021	&	Traditional, IoT	&	BoT-IoT, UNSW-NB15	&	White-box	&	Evasion	&	C\&W, FGSM, JSMA	&	DT, RF, SVM	&	N/A	\\ \cline{2-10}
&	~\cite{debicha2021adversarial}	&	2021	&	Traditional	&	NSL-KDD	&	White-box	&	Evasion	&	BIM, FGSM, PGD	&	DNN	&	Adversarial Training	\\ \cline{2-10}
&	~\cite{fu2021robust}	&	2021	&	Traditional, IoT	&	CICIDS2018	&	White-box	&	Evasion	&	FGSM	&	CNN, LSTM, Gated Recurrent Unit	&	Adversarial Training	\\ \cline{2-10}
&	~\cite{maarouf2021evaluating}	&	2021	&	Traditional	&	SCX VPN-NonVPN, NIMS	&	White-box, Black-box	&	Evasion	&	Deepfool, PGD, Zoo	&	CNN, DNN, K-NN, Recurrent NN, C4.5	&	N/A	\\ \cline{2-10}
	&	~\cite{piplai2020nattack}	 & 	2020	 & 	Traditional 	 & 	IEEE Big Data2019	 &	White-box	 & 	Evasion	 & 	FGSM 	 & 	GAN	 & 	Adversarial Training	\\ \hline
Syn.	&	~\cite{zhang2019deep1}	 & 	2019	 & 	Traditional 	 & 	KDDCUP99	 &	N/A	 & 	N/A	 & 	PGJPG, MC, GAN	 & 	DNN, LR, SVM	 & 	N/A	\\ \hline

\caption{Summary of surveyed studies\\
\textbf{Def.} Defend ML-based NIDS Models’ against AEs\\ \textbf{Eva.} Evaluate ML-based NIDS Models’ Resilience against AEs\\ \textbf{Gen.} Generate AEs to Attack ML-based NIDS Models\\ \textbf{Syn.} Synthesize Malicious Network Traffic}
\label{table:Summary of surveyed studies}
\end{longtable}
\twocolumn
\normalsize
\bibliographystyle{IEEEtranN}
\footnotesize
\bibliography{Bibliography.bib}
\end{document}